# Size Effect of Soft Phonon Dispersion in Nanosized Ferroics

Anna N. Morozovska[1,2*] and Eugene A. Eliseev[3†]

[1] *Institute of Physics, National Academy of Sciences of Ukraine,*

*46, Prospekt Nauky, 03028 Kyiv, Ukraine*

[2] *Bogolyubov Institute for Theoretical Physics, National Academy of Sciences of Ukraine,*

*14-b Metrolohichna, 03680 Kyiv, Ukraine*

[3] *Institute for Problems of Materials Science, National Academy of Sciences of Ukraine,*

*3, Krjijanovskogo, 03142 Kyiv, Ukraine*

**Abstract**

Using Landau-Ginsburg-Devonshire theory, we derive and analyze analytical expressions for the frequency dispersion of soft phonon modes in nanosized ferroics and perform numerical calculations for a thin $SrTiO_3$ film. We revealed the pronounced "true" size effect in the dependence of soft phonon spatial dispersion on the film thickness and predict that it can lead to the "apparent" or "false" size effect of dynamic flexoelectric coupling constants. Also we derived analytical expressions describing the influence of finite size effect on the appearance and properties of the incommensurate spatial modulation induced by the static flexoelectric effect in ferroic thin films. To verify the theoretical predictions experimental measurements of the soft phonon dispersion in nanosized ferroics seem urgent.

---

[*] anna.n.morozovska@gmail.com
[†] eugene.a.eliseev@gmail.com

# I. INTRODUCTION

Microstructural, electrophysical, electromechanical and magnetoelectric properties of nanosized ferroics (shortly "**nanoferroics**"), such as thin films and nanoparticles of proper and incipient ferroelectric, quantum paraelectrics, ferroelastics and ferromagnetics with inherent phase transitions to the long-range ordered phase(s), attract permanent attention of researchers [1, 2]. In particular, experimental and theoretical studies of the lattice dynamics in nanoferroics are the inexhaustible source of versatile valuable information for fundamental physical research and advanced applications [1-2]. Notably that any phase transition leads to the instability of the definite phonon vibration modes, at that the static displacements of atoms at the phase transition correspond to frozen displacements of the soft phonon modes [3, 4]. Basic experimental methods, which contain information about the soft phonon modes and spatial modulation of the order parameter in ferroics are dielectric measurements [5], inelastic neutron scattering [3, 6, 7, 8, 9, 10], X-ray [11, 12, 13], Raman [14] and Brillouin [11, 12, 15, 16, 17, 18] scatterings and ultrasonic pulse-echo method [15, 17] allowing hypersound spectroscopic measurements.

The flexoelectric coupling is of great importance for nanoscale objects [19, 20] and especially for nanoferroics [21, 22], where the strong strain gradients are inevitably present at the surfaces of thin films [23, 24, 25], domain walls and interfaces [26, 27, 28]. The static flexoelectric effect, firstly introduced by Mashkevich, Tolpygo [29], and Kogan [30], manifesting itself in the appearance of electric polarization variation $\delta P_i$ linearly proportional to the strain gradient $\partial u_{kl}/\partial x_j$, exists in all solids, making the effect universal [31, 32, 33]. The proportionality coefficient $f_{ijkl}$ in the linear relation, $\delta P_i = f_{ijkl} \frac{\partial u_{kl}}{\partial x_j}$, is the component of the flexoelectric coupling tensor. Relatively recently Kvasov and Tagantsev [34] predicted the existence of dynamic flexoelectric effect contributing to the polarization variation as $\delta P_i = -\frac{M_{ij}}{\alpha} \frac{\partial^2 U_j}{\partial t^2}$, where $U_j$ are the components of elastic displacement, $M_{ij}$ are the components of the dynamic flexoelectric tensor and α is the dielectric stiffness.

Considering the importance of the flexoelectric coupling (shortly "**flexocoupling**") for the physical understanding of the gradient-driven couplings in mesoscale and nanoscale solids, one has to determine its symmetry and numerical values. Unfortunately, the available experimental and theoretical data about the flexocoupling tensor symmetry, specifically the amount of independent components allowing for the point group symmetry [35, 36] and "hidden" permutation symmetry [37], and numerical values are contradictory [38]. Namely, the upper limits for the values $f_{ijkl}$ established by Yudin et al [39], as well as the values calculated from the first principles for bulk ferroics [40, 41, 42, 43, 44], the can be several orders of magnitude smaller than those measured experimentally in ferroelectric ceramics [45, 46, 47] and thin films [48], ferroelectric relaxor polymers [49] and electrets [50], incipient ferroelectrics [51, 52] and biological membranes [53, 54]. Bersuker argued [55] that the anomalously high flexoelectric coefficients in perovskite ceramics may be related with the manifestation of the pseudo Jahn-Teller effect. Steaming from a vibronic nature, pseudo Jahn-Teller effect can affect the dynamic flexoeffect, and, indeed, the available information about the numerical values of $M_{ij}$ is completely controversial, because, on one hand, there are microscopic theories in which the effect is absent [43], and, on the other hand, its determination from the soft phonon spectra leads to nonzero $M_{ij}$ [34, 56, 57], which impact appeared comparable to that of the static flexoelectric effect in e.g. bulk $SrTiO_3$.

The impact of static and dynamic flexoelectric effect on the dispersion law $\omega(k)$ of the soft phonon frequency ω in ferroics has been studied recently in Refs.[56, 57], where we used Landau-Ginzburg-Devonshire (**LGD**) approach to model the properties of optic and acoustic phonons in the ferroelectric and paraelectric phases of different ferroics. It appeared that the joint action of static and dynamic flexoelectric effect essentially broadens the k-spectrum of generalized susceptibility and leads to the additional "pushing away" of the optical and acoustic soft mode phonons. The degeneration of the transverse optic and acoustic modes disappears in the ferroelectric phase in comparison with the paraelectric phase due to the synergy of flexoelectric coupling and ferroelectric nonlinearity. The soft acoustic modes and spatially modulated phases (**SPM**) can be induced by the flexoelectric coupling in ferroics [22, 58, 59] and the surface acoustic waves can propagate in non-piezoelectric solids due to the static and dynamic flexoelectric effects [60].

Let us underline that the theoretical results [56-60], which can be important for the theoretical analyses of the neutron, Raman and Brillouin scattering experiments [22], have been obtained for unconfined [56-59] or semi-infinite [60] ferroics. However, to the best of our knowledge, the theoretical description of the phonon dispersion in nanoferroics (e.g. in thin films or nanoparticles) is still absent. Therefore in this study we derive and analyze the dispersion $\omega(k)$ of soft phonon modes in dependence on the thickness of ferroic film allowing for the static and dynamic flexocoupling; and perform numerical calculations for an incipient ferroelectric $SrTiO_3$ using LGD theory. The paper is structured in the following way. The problem formulation, used approximations and analytical solution for the phonon dispersion in a thin paraelectric film are presented in **Section II**. Obtained analytical expressions for $\omega(k)$ are discussed and illustrated on example of $SrTiO_3$ thin film in **Section III**, where the special attention is paid to the "true" size effect of the soft phonon dispersion and to the "apparent" or "false" size effect of the dynamic flexocoupling, as well as to the influence of the film thickness on the SPM appearance and critical values of the static flexocoupling constants. **Section IV** is a brief summary. The details of calculations are presented in the **Supplement** [61].

## II. ANALYTICAL SOLUTION FOR A THIN FERROELECTRIC FILM
### A. The problem formulation

Let us consider a ferroic film of thickness $2h$ confined in $z$-direction. The film is placed between two ideally conducting electrodes and clamped to a matched substrate that does not produce a misfit strain in the film. The only nonzero components of the long-range order parameters are y-components of electric polarization $P_y(x,z)$, further denoted as $\eta(x,z)$, and mechanical displacement $U_y(x,z)$, further denoted as $U(x,z)$. The geometry of the problem is shown in **Fig.1**.

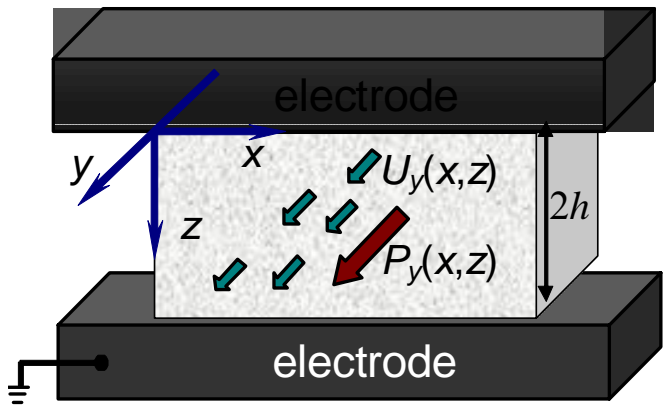

**FIG.1.** The geometry of the ferroic film confined in z-direction. The y-components of electric polarization $P_y(x,z)$ and related component of mechanical displacement $U_y(x,z)$ are nonzero.

In contrast to the three-component case [56], the one-component approximation for polarization and strain allows considering only the lowest transverse optic and acoustic soft phonon modes in ferroics [57-59], because $div\vec{P} = 0$ and $div\vec{U} = 0$ in the case. As it was shown in Refs.[57, 58], the one-component approximation can describe quantitatively the experimental data for the soft phonon dispersion in paraelectric SrTiO$_3$ [6], uniaxial ferroelecrics Sn$_2$P$_2$S$_6$ [62] and Sn$_2$P$_2$Se$_6$ [63], and, quite unexpectedly, the dispersion of the lowest modes in the paraelectric phase of ferroelectric PbTiO$_3$ [3] and organic ferroelectric (CH$_3$)$_3$NCH$_2$COO·CaCl$_2$·2H$_2$O [9]. However, strictly speaking, the one-component approximation cannot describe the soft phonon dispersion in multiaxial ferroelectrics, such as perovskites BaTiO$_3$ or (Pb,Zr)TiO$_3$, wherein three acoustic and several optic phonon modes are observed [56]. This is because the simplified dispersion law, obtained within the approximation, cannot describe the interaction between different transverse and longitudinal optic and three acoustic modes induced by cooperative effects, flexoelectric and electrostriction couplings in multiaxial ferroelectrics (see e.g., the models by Kappler, Walker [64], and Hlinka et al [65]). Unfortunately, it appeared impossible for us to derive analytical expressions for phonon dispersion in thin films in a general three-dimensional case, and, for the sake of clarity, we were limited ourselves by the one-dimensional approximation.

Within the one-dimensional approximation, dynamic equations for the order parameters $\eta$ and $u$ can be obtained by the variation of the Lagrange function of the film that has the following form:

$$L_V = \int_0^\infty dt \int_{-\infty}^\infty dx \int_{-h}^h dz (F_V - K_V), \tag{1a}$$

$$F_V = \frac{\alpha(T)}{2}\eta^2 + \frac{\beta}{4}\eta^4 + \frac{\gamma}{4}\eta^6 - \eta E + \frac{g_l}{2}\left(\frac{\partial \eta}{\partial z}\right)^2 + \frac{g_\perp}{2}\left(\frac{\partial \eta}{\partial x}\right)^2 - \frac{f_l}{2}\left(\eta \frac{\partial^2 U}{\partial z^2} - \frac{\partial U}{\partial z}\frac{\partial \eta}{\partial z}\right)$$
$$- \frac{f_\perp}{2}\left(\eta \frac{\partial^2 U}{\partial x^2} - \frac{\partial U}{\partial x}\frac{\partial \eta}{\partial x}\right) - q\frac{\partial U}{\partial z}\eta^2 + \frac{c_l}{2}\left(\frac{\partial U}{\partial z}\right)^2 + \frac{c_\perp}{2}\left(\frac{\partial U}{\partial x}\right)^2 + \frac{v}{2}\left(\frac{\partial^2 U}{\partial z^2}\right)^2 + N\frac{\partial U}{\partial z}, \tag{1b}$$

$$K_V = \left(\frac{\mu}{2}\left(\frac{\partial \eta}{\partial t}\right)^2 + M\frac{\partial \eta}{\partial t}\frac{\partial U}{\partial t} + \frac{\rho}{2}\left(\frac{\partial U}{\partial t}\right)^2\right). \tag{1c}$$

The surface free energy related with the polar order-parameter is

$$F_S = \frac{\zeta}{2}\int_{-\infty}^\infty dx\left[\eta^2(x,-h) + \eta^2(x,h)\right] \tag{2}$$

According to Landau theory [66, 67], the coefficient $\alpha$ in Eq.(1b) linearly depends on the temperature $T$ for proper ferroelectrics, $\alpha(T) = \alpha_T(T - T_C)$, where the $\alpha_T > 0$ and $T_C > 0$ is the Curie temperature. For incipient ferroelectrics (such as SrTiO$_3$) $\alpha$ obeys the Barrett-type formula [68], $\alpha(T) = \alpha_T(T_q \coth(T_q/T) - T_C)$, where $T_C \leq 0$ is the "virtual" Curie temperature, $T_q > 0$ is a characteristic temperature. $q$ is the longitudinal component of electrostriction tensor (the transverse component is regarded zero for chosen geometry), $f_l$ are $f_\perp$ the longitudinal and transverse components of the static flexoelectric tensor, respectively [69]. The higher order coefficient $\beta$ (that can be positive or negative depending on the ferroic order) and positive coefficient $\gamma$ are regarded temperature independent; $g_l$ and $g_\perp$ are positive longitudinal and transverse components of the polarization gradient tensor, $c_l$ and $c_\perp$ are elastic compliances in the normal and transverse directions with respect to the film surfaces, $z = -h$ and $z = h$, respectively. The longitudinal coefficient $v$ of higher elastic gradient is not negative, $v \geq 0$, and its transverse component does not contribute for chosen geometry. $M$ is the dynamic flexocoupling coefficient; $\rho$ is the ferroic mass density at normal conditions, and $\mu$ is the kinetic coefficient in kinetic energy (1c). The coefficient $\zeta \geq 0$ in the surface energy (2) is weakly temperature dependent. Also we can apply the electric field E and/or elastic force N to the ferroic film.

Euler-Lagrange equations for $\eta(x,z)$ and $U(x,z)$, obtained by varying Eq.(1a), are listed in **Supplement.** The boundary conditions for $\eta(x,z)$ derived by varying Eqs.(1) on $\eta$ have the form:

$$\left. -\zeta\eta - g_l \frac{\partial \eta}{\partial z} - \frac{f_l}{2}\left(\frac{\partial U}{\partial z}\right) \right|_{z=h} = 0, \quad \left. \zeta\eta - g_l \frac{\partial \eta}{\partial z} - \frac{f_l}{2}\left(\frac{\partial U}{\partial z}\right) \right|_{z=-h} = 0. \quad (3)$$

Supplementary boundary conditions for $U(x,z)$ make sense only in the case of positive $v$, and they are listed in **Supplement** for the considered geometry. To obtain and analyze relatively simple analytical expressions, one can show that the conditions for $U(x,z)$ can be neglected in the case of small enough $v$ and under the validity of strict equality $f^2 \ll cg$ regarded valid hereinafter (for details see Ref.[57] and the criteria of upper bonds of flexoelectric coefficients in Ref.[39]). The higher gradient term $\frac{v}{2}\left(\frac{\partial^2 U}{\partial z^2}\right)^2$ is usually omitted, except several papers, e.g., Eliseev et al [70], Yurkov [71], Mao and Purohit [72], and Stengel [44]. Notably, the supplementary conditions for $U(x,z)$ become redundant in the limit $v=0$ considered hereinafter.

Let us seek for the solution of the problem for the paraelectrics phase, where $\alpha(T) > 0$ and the spontaneous order parameter is zero. The expansion of $\eta(x,z)$ and $U(x,z)$ in Fourier integral in k-$\omega$ space have the form:

$$\eta = \int dk_\perp dk_l \exp(ik_\perp x + ik_l z + i\omega t)\tilde{\eta}(\vec{k}), \quad U = \int dk_\perp dk_l \exp(ik_\perp x + ik_l z + i\omega t)\tilde{U}(\vec{k}) \quad (4)$$

Linearized Euler-Lagrange equations in k-$\omega$ space acquire the following form

$$\left(-\mu\omega^2 + \alpha + g_\perp k_\perp^2 + g_l k_l^2\right)\tilde{\eta} + \left(f_l k_l^2 + f_\perp k_\perp^2 - M\omega^2\right)\tilde{U} = \tilde{E}, \quad (5a)$$

$$\left(c_\perp k_\perp^2 + c_l k_l^2 - \rho\omega^2\right)\tilde{U} + \left(f_l k_l^2 + f_\perp k_\perp^2 - M\omega^2\right)\tilde{\eta} = -ik_l \tilde{N}. \quad (5b)$$

As a result the Fourier images of the order parameter $\tilde{\eta}(\vec{k})$ and elastic displacement $\tilde{U}(\vec{k})$ are linearly proportional to external electric field and mechanical force variations, $\tilde{E}$ and $\tilde{N}$. The formal solution of Eqs.(5) can be presented in the matrix form and are listed in **Supplement.**

The condition when the determinant of the system (5) is zero lead to the expression for the phonon frequency dispersion, $\omega(k)$, in a bulk ferroic,

$$\omega_{1,2}^2(\vec{k}) = \frac{C(\vec{k}) \pm \sqrt{C^2(\vec{k}) - 4(\mu\rho - M^2)B(\vec{k})}}{2(\mu\rho - M^2)}, \quad (6)$$

where the functions $C(\vec{k}) = \alpha\rho + (\mu c_\perp - 2f_\perp M + \rho g_\perp)k_\perp^2 + (c_l\mu - 2f_l M + g\rho)k_l^2$ and $B(\vec{k}) = (\alpha + g_l k_l^2 + g_\perp k_\perp^2)(c_l k_l^2 + c_\perp k_\perp^2) - (f_l k_l^2 + f_\perp k_\perp^2)^2$, respectively (see **Supplement** for details). Dispersion relation (6) contains the transverse optical (**O**) and acoustic (**A**) phonon modes, which corresponds to the signs "+" and "−" before the radical, respectively.

### B. Approximate analytical solution for thin films

For thin films the boundary conditions (3) makes the $k_l$-spectra of $\omega(\vec{k})$ discrete, $k_l \to k_m$, where $m$ are integer numbers. The discrete eigen values $k_m$ were determined from zero determinant $\|D_m\|$ of the system (5) along with the boundary conditions (3) (see **Supplement**). It turned out, that the condition $\|D_m\| = 0$ is equivalent to the characteristic equation

$$\text{Im}[A(k_m,\omega)]\cos(2k_m h) + \text{Re}[A(k_m,\omega)]\sin(2k_m h) = 0, \tag{7}$$

where

$$A(k_m,\omega) = -\left(\zeta - ik_m g_l - ik_m \frac{f_l}{2}\frac{f_l k_m^2 + f_\perp k_\perp^2 - M\omega^2}{c_l k_m^2 + c_\perp k_\perp^2 - \rho\omega^2}\right)^2. \tag{8}$$

The soft phonons dispersion $\omega(k_m, k_\perp)$ is given by Eq.(6), but with discrete $k_m$, namely:

$$\omega_{1,2}^2(k_m, k_\perp) = \frac{C(k_m, k_\perp) \pm \sqrt{C^2(k_m, k_\perp) - 4(\mu\rho - M^2)B(k_m, k_\perp)}}{2(\mu\rho - M^2)}, \tag{9}$$

where the functions $C(k_m, k_\perp) = \alpha\rho + (\mu c_\perp - 2f_\perp M + \rho g_\perp)k_\perp^2 + (c_l\mu - 2f_l M + g\rho)k_m^2$ and $B(k_m, k_\perp) = (\alpha + g_l k_m^2 + g_\perp k_\perp^2)(c_l k_m^2 + c_\perp k_\perp^2) - (f_l k_m^2 + f_\perp k_\perp^2)^2$ also depend on the film thickness $h$ and parameter $\zeta$ in the boundary conditions (3) due to the dependences of $k_m$ on $h$ and $\zeta$ per Eq.(7).

### C. Limiting cases of the boundary conditions

The limiting case of zero derivative of the order parameter and strain at the film surfaces, $\left.\frac{\partial\eta}{\partial z}\right|_{z=\pm h} = 0$ and $\left.\frac{\partial U}{\partial z}\right|_{z=\pm h} = 0$, corresponds to $\zeta \to 0$ and leads to the expression $A(k_m,\omega) = \left(g_l + \frac{f_l}{2}\frac{f_l k_m^2 + f_\perp k_\perp^2 - M\omega^2}{c_l k_m^2 + c_\perp k_\perp^2 - \rho\omega^2}\right)^2 k_m^2$ in Eq.(8). Since $A(k_m,\omega) \geq 0$ for $\zeta \to 0$ and the variable $k_\perp^2$ is independent, Eq.(7) reduces to the equation $\sin(2k_m h) = 0$, that solution has the form $k_m = \frac{\pi m}{2h}$, where $m = 0, 1, 2,\ldots$ However for the case one should consider the lowest root $m=0$ corresponding to the constant order parameter, $\eta = \eta_S$, and obtain the bulk expressions (6) for the soft phonon spectra.

For the limiting case of the zero order parameter at the film surfaces, $\eta|_{z=\pm h} = 0$, that is the most favorable case for the observation of finite size effects, the constant $\zeta \to \infty$ in the boundary conditions (3). Since $A(k_m, \omega) \to -\zeta^2$ for $\zeta \to \infty$, the condition $\sin(2k_m h) = 0$ should be valid for the validity of Eq.(7). The equation $\sin(2k_m h) = 0$ gives a well-known solution for eigen wave numbers $k_m = \frac{\pi m}{2h}$, where $m = 0, 1, 2, \ldots$. Since $m=0$ corresponds to identically zero solution for the order parameter, one should consider $m=1$ corresponding to the lowest root. For the case the first harmonic in the order parameter distribution is

$$\eta = \int dk_\perp \exp(ik_\perp x + i\omega t)\tilde{\eta}(k_\perp, k_l)\cos\left[\frac{\pi z}{2h}\right], \quad k_l = \frac{\pi}{2h}, \quad (10)$$

The soft phonons dispersion $\omega(h, \vec{k})$ given by Eq.(9) acquires the form:

$$\omega_{1,2}^2(h, k_\perp) = \frac{C(h, k_\perp) \pm \sqrt{C^2(h, k_\perp) - 4(\mu\rho - M^2)B(h, k_\perp)}}{2(\mu\rho - M^2)}, \quad (11)$$

where the functions $C(h, k_\perp) \approx \alpha(T)\rho + (\mu c_\perp - 2f_\perp M + \rho g_\perp)k_\perp^2 + (c_l\mu - 2f_l M + g_l\rho)(\pi/2h)^2$ and $B(h, k_\perp) \approx (\alpha(T) + g_l(\pi/2h)^2 + g_\perp k_\perp^2)(c_l(\pi/2h)^2 + c_\perp k_\perp^2) - (f_l(\pi/2h)^2 + f_\perp k_\perp^2)^2$ depend on the film thickness $h$. Different signs "−" and "+" correspond to the frequencies $\omega_1(h, k_\perp)$ and $\omega_2(h, k_\perp)$, respectively.

### III. RESULTS AND DISCUSSION
#### D. Size effect of the soft phonon dispersion

**Figure 2(a)** illustrates the dependences of the phonon frequency $\omega$ on the wave vector $k$ calculated for SrTiO$_3$ thin films with different thickness $h = (6 - 60)$ lattice constants (*l.c.*) (colored solid curves) in comparison with a thick film (black dashed curve) calculated from Eq.(11) at room temperature. The frequency $\omega_1(h, k_\perp)$ monotonically decreases with $h$ increase, but it is nonzero at $k_\perp = 0$ for any thickness and tends to the expression (6) for the "bulk" A-mode $\omega_1(\infty, k_\perp \to 0) \to k_\perp\sqrt{c/\rho}$ only at $h \to \infty$. This is a reminiscent difference between the phonon spectrum in a thin film with discrete $k_z$ values and the spectrum of a bulk material with continuous $k_z$ values. In particular for finite thickness $h$ and $k_\perp = 0$ one obtains from Eq.(11) that

$$\omega_1^2(h,0) \approx \frac{c_l}{\rho}\left(\frac{\pi}{2h}\right)^2 + \frac{g_l c_l - f_l^2}{\alpha\rho}\left(\frac{\pi}{2h}\right)^4 \qquad (12a)$$

Approximate equality in Eq.(12) is valid for enough high thicknesses $h \gg a$ ($a$ is a lattice constant) and $\alpha > 0$ [see Eq.(A.29a) in **Supplement**]. So that $\omega_1(h \gg a, k_\perp \to 0) \sim \sqrt{\frac{c_\perp}{\rho}k_\perp^2 + \frac{c_l}{\rho}\left(\frac{\pi}{2h}\right)^2}$ as anticipated.

As is obvious from the figure, the frequency $\omega_2(h, k_\perp)$ also monotonically decreases with $h$ increase and tends to expression (6) for the "bulk" O-mode $\omega_1(\infty, k_\perp)$ for $h>30$ *l.c.*. For $k_\perp = 0$ the value $\omega_2$ significantly depends on the thickness $h$ until $h<30$ *l.c.*, namely one obtains from Eq.(11) that

$$\omega_2^2(h,0) \approx \frac{\alpha}{\mu} + \left(\frac{c_l}{\mu} - \frac{2f_l M}{\mu\rho} + \frac{g_l}{\mu}\right)\left(\frac{\pi}{2h}\right)^2 \qquad (12b)$$

Approximate equality in Eq.(12b) is valid in the case $h \gg a$ and $\mu\rho \gg M^2$ that is typical for paraelectrics (e.g. for SrTiO$_3$).

**Figure 2(b)** illustrates the thickness dependences of $\omega_1$ (blue curve), $\omega_2$ (red curve) and their difference $\omega_2 - \omega_1$ (black curve) calculated at $k_\perp = 0$ and 300 K. A dashed vertical line indicates the estimate for the thickness limit (~ 5 – 10 *l.c.*) of LGD approach applicability. Both $\omega_1$ and $\omega_2$ rapidly increases with $2h$ decrease below 20 nm and tend to the thickness-independent constants at $2h > 50$ nm. The pronounced size effect (i.e. the minimum) is seen at the dependence of $\omega_2 - \omega_1$ on the film thickness varying in the range (6 – 20) nm. However it is questionable whether the prediction can be verified experimentally, because it can be rather difficult to determine experimentally (e.g. by neutron scattering) the phonon dispersion at very small **k** in ultrathin SrTiO$_3$ films. However the Brillouin or Raman scattering of nanoferroics probably can give the valuable information about the dispersion law $\omega(h,k)$.

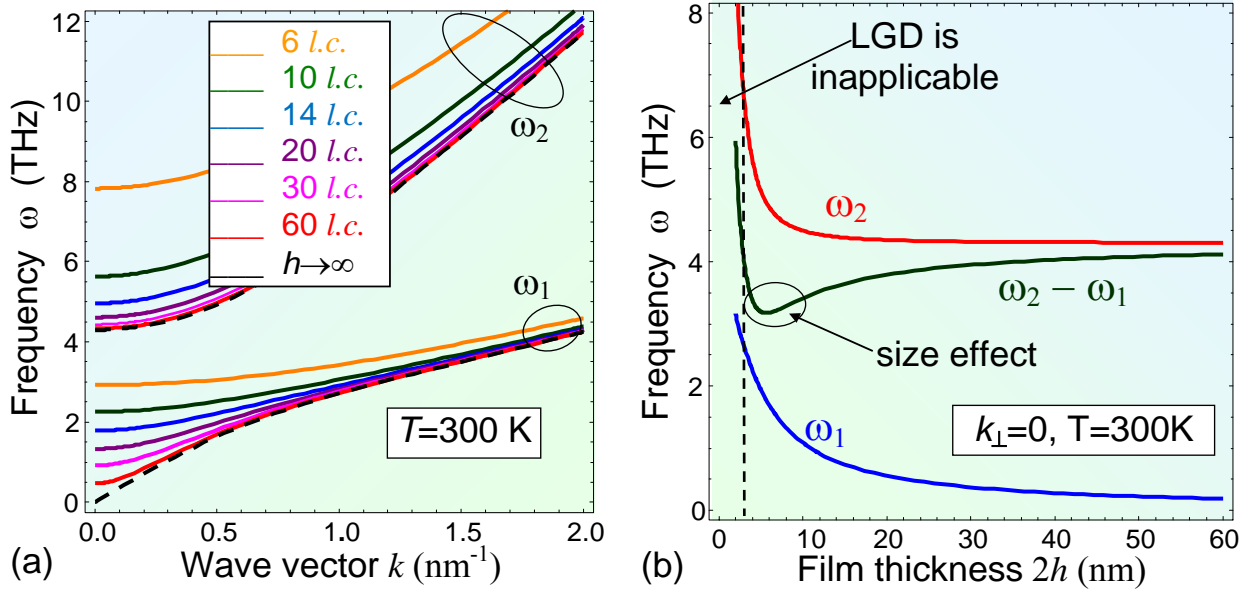

**FIG. 2. (a)** Dependence of the phonon frequency $\omega$ on the wave vector $k_\perp$ calculated for SrTiO$_3$ thin films with different thickness $2h$= 6, 10, 14, 20, 30, 60 $l.c.$ (curves 1-6) and bulk SrTiO$_3$ (curve 7) at temperature 300 K. **(b)** Thickness dependences of $\omega_1$ (blue curve), $\omega_2$ (red curve) and $\omega_2 - \omega_1$ (black curve) calculated at $k_\perp = 0$ and 300 K. The dashed vertical line indicates the thickness limit (~ 6 lattice constants) of LGD approach applicability. Parameters of SrTiO$_3$ used in the calculation are listed in **Table I**.

**TABLE I.** Description of the symbols in the Lagrange function (1), and their numerical values for SrTiO$_3$

| Description | Symbol and dimension | Incipient ferroelectric SrTiO$_3$ |
|---|---|---|
| Coefficient at $\eta^2$ | $\alpha(T)$ ($\times$C$^{-2}\cdot$mJ) | $\alpha_T(T_q \coth(T_q/T) - T_C)$ |
| Curie-Weiss constant | $\alpha_T$ ($\times 10^5$C$^{-2}\cdot$mJ/K) | 15 |
| Curie temperature | $T_C$ (K) | $T_C$=30, $T_q$=54 |
| LGD-coefficient at $\eta^4$ | $\beta$ ($\times 10^8$ JC$^{-4}\cdot$m$^5$) | 81 |
| LGD-coefficient at $\eta^6$ | $\gamma$ ($\times 10^9$JC$^{-6}\cdot$m$^9$) | 0 (data is absent) |
| Electrostriction | $q$ ($\times 10^9$Jm/C$^2$) | $q_{44}$=2.4 * |
| Elastic stiffness | $c$ ($\times 10^{10}$ Pa) | $c_l=c_\perp=c_{44}$=12.7 |
| Gradient coefficient at $(\nabla \eta)^2$ | $g$ ($\times 10^{-10}$C$^{-2}$m$^3$J) | $g_l=g_\perp=g_{44}$=2 |
| Strain gradient $(\nabla u)^2$ | $v$ ($\times 10^{-9}$V s$^2$/m$^2$) | 0 (data is absent) |
| Static flexocoupling | $f$ (V) | $f_l=f_\perp=f_{44}$=2.2 |
| Dynamic flexocoupling | $M$ ($\times 10^{-8}$Vs$^2$/m$^2$) | 2 |
| Kinetic coefficient | $\mu$ ($\times 10^{-18}$s$^2$mJ) | 22 |
| Material density at n. c. | $\rho$ ($\times 10^3$ kg/m$^3$) | 4.930 |
| Lattice constant ($l.c.$) | $a$ (nm) | 0.395 |

* the parameter is not used in our calculations

### E. "Apparent" or "false" size effect of the dynamic flexoelectric coupling

Let us underline that the available information about numerical values of the static and especially dynamic flexoelectric tensor components is still controversial (since calculations and different experiments give different results) or even unknown [38-50], the analysis of the soft phonon spectra can be one of the most reliable way to define it. Notably that expressions (11) allows us to define the relationship between the "true" static ($f_l$) and "effective" dynamic ($M$) flexoelectric constants (shortly "**flexoconstants**") in thin ferroelectric films from the interpolation of the soft phonon dispersion $\omega_{1,2}(h, k_\perp)$ in the limit $k_\perp \to 0$. Actually, elementary transformations of Eqs.(11) lead to the expression for the constant $M$:

$$M_{1,2}[h,\omega] = f_l \left(\frac{\pi}{2h\omega}\right)^2 \pm \sqrt{f_l^2 \left(\frac{\pi}{2h\omega}\right)^4 - \left(\frac{\pi}{2h\omega}\right)^2 (c_l \mu + g_l \rho) + \rho\left(\mu - \frac{\alpha}{\omega^2}\right)}. \quad (13a)$$

Hereinafter $\omega^2 \equiv \omega_1^2(h,0) + \omega_2^2(h,0)$. Derivation of Eq.(13a) in given in the end of **Supplement.** The approximate expressions can be derived from Eq.(13a) in a deep paraelectric phase ($\alpha > 0$) for the case $\mu\rho \gg M^2$ typical for all known incipient and proper ferroelectric perovskites:

$$M_{1,2}[h,\omega] \approx \begin{cases} \dfrac{1}{2f_l}\left[\rho(\alpha - \omega^2 \mu)\left(\dfrac{2h}{\pi}\right)^2 + c_l\mu + g_l\rho\right], & h \to 0 \\ \pm\sqrt{\rho\left(\mu - \dfrac{\alpha}{\omega^2}\right)}, & h \to \infty \end{cases} \quad (13b)$$

As it follows from Eq.(13b) that first line contains the evident dependence of $M$ on $h$, $M \sim \dfrac{1}{f_l}\left(\dfrac{2h}{\pi}\right)^2$, the determination of $\omega^2(h,0)$ from experiments performed for a set of small $h$ may lead to the conclusion about the existence of the "apparent" or "false" size effect of the dynamic flexocoupling at fixed other parameters.

The apparent size effect of dynamic flexocoupling is illustrated in **Fig. 3(a)** at fixed frequency $\omega^2 = const$ and room temperature**.** The flexoconstants $M_1$ (shown by red curve) and $M_2$ (shown by blue curve) significantly increase with $2h$ increase from 14 nm to 30 nm and gradually tend to the "bulk" values $M_{1,2}[\infty, \omega] = \pm\sqrt{\rho\left(\mu - \dfrac{\alpha}{\omega^2}\right)}$ at $2h > 35$ nm. At the same time their sum $M_1 + M_2$ (shown by magenta curve) significantly decreases with $2h$ increase from 14 nm to 30 nm and then tends to zero at $2h > 35$ nm. The real physical roots $M_{1,2}[h,\omega]$ exist only for $h > h_{cr}$, where the critical thickness $h_{cr}$ depends on the frequency $\omega$ and is about 14 *l.c.* for $\omega = 4.5$ THz.

**Fig.3(b)** shows the dependences of $M_1$, $M_2$ and $M_1 + M_2$ on $\omega$ at fixed thickness $2h=8$ nm. The constants $M_1$ (shown by red curve) and $M_2$ (shown by blue curve) significantly increase with $\omega$ increase from 4.9 THz to 9 THz, and then gradually tend to the "bulk" values $M_{1,2}[\infty, \omega]$ at $\omega > 9$ THz. At the same time their sum $M_1 + M_2$ (shown by magenta curve) significantly decreases with $\omega$ increase from 4.9 THz to 9 THz, and then gradually tend to zero at $\omega > 9$ THz. The real physical roots $M_{1,2}[h, \omega]$ exist only for $\omega > \omega_{cr}$, where the critical frequency $\omega_{cr}$ depends on $h$ and is about 4.9 THz for $h=8$ nm.

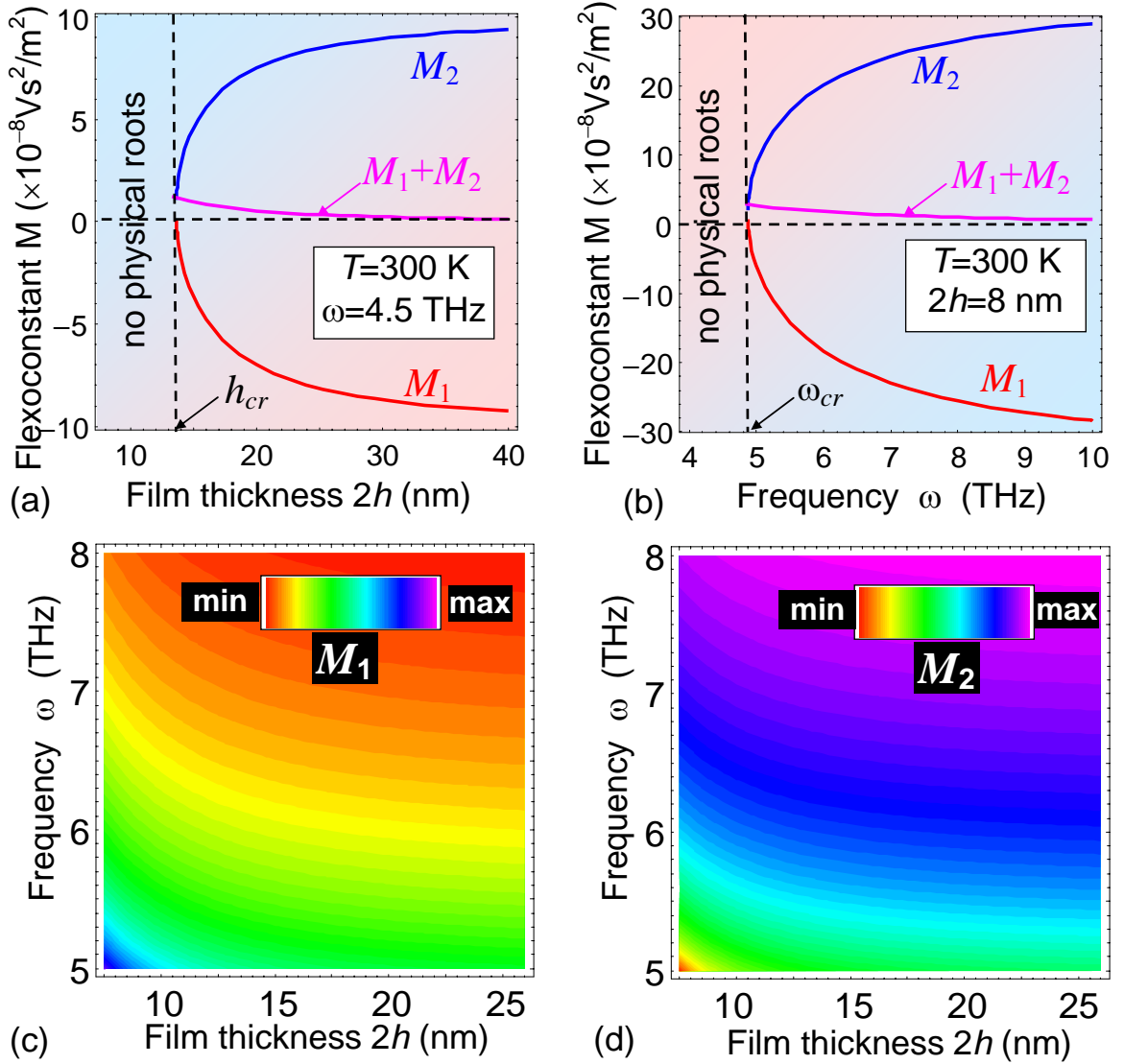

**FIG. 3.** (a) The "apparent" dependences of the dynamic flexocoupling constants $M_1$ (red curve), $M_2$ (blue curve) and their sum $M_1 + M_2$ (magenta curve) on SrTiO$_3$ film thickness $2h$ calculated from Eq.(13a) at frequency $\omega = 4.5$ THz and $T=300$ K. (b) The dependences of $M_1$ (red curve), $M_2$ (blue curve) and $M_1 + M_2$ (magenta curve) on $\omega$ calculated from Eq.(13a) for SrTiO$_3$ film thickness $2h=8$ nm. Contour maps

of the roots $M_1$ **(c)** and $M_2$ **(d)** calculated in coordinates ω and h. Other parameters of SrTiO$_3$ used in the calculation are listed in **Table I**.

Contour maps of the roots $M_1$ and $M_2$ plotted in coordinates ω and h are shown in **Figs.3(c)** and **3(d)**, respectively. From the maps the apparent size effect exists in the thickness range less than 25 nm. We would like to underline that the "apparent" size effect of dynamic flexocoupling coefficients [shown in **Figs.3**] should be clearly distinguished from the "true" size effect of the soft phonon frequency, predicted by us and demonstrated on example of SrTiO$_3$ in **Figs.2.**

On the other hand it is not excluded that the origin of the "giant" flexoelectric coefficients extracted by several authors based on experimental results for a spatially confined or strongly inhomogeneous samples is the apparent size effect of dynamic constant. Qualitatively, the dynamic flexoeffect may arise from pseudo Jahn-Teller vibrations [55].

### F. Size effect of the spatially modulated structures

The SMP with the modulation period $2\pi/k_\perp$ can appear under the condition of the static dielectric susceptibility divergence [56 - 58]. For thin film the expression for the static susceptibility $\widetilde{\chi}_S(h, k_\perp)$ is derived in **Supplement** and has the form:

$$\frac{1}{\widetilde{\chi}_S(h, k_\perp)} = \alpha(T) + g_l \left(\frac{\pi}{2h}\right)^2 + g_\perp k_\perp^2 - \frac{\left(f_l(\pi/2h)^2 + f_\perp k_\perp^2\right)^2}{c_l(\pi/2h)^2 + c_\perp k_\perp^2} \to 0 \quad (14)$$

Equation (14) is equivalent to the conditions $\omega_1(h, k_\perp) = 0 \leftrightarrow B(h, k_\perp) = 0$ in Eq.(11). The condition (14) leads to the biquadratic equation for $k_\perp$, which solutions are

$$k_\perp^2 = \frac{B - \sqrt{B^2 - 4AC}}{2A}, \quad \text{and} \quad k_\perp^2 = \frac{B + \sqrt{B^2 - 4AC}}{2A}. \quad (15)$$

Where the parameters $A = c_\perp g_\perp - f_\perp^2$, $B = c_\perp + (c_\perp g_l + c_l g_\perp - 2 f_l f_\perp)\left(\frac{\pi}{2h}\right)^2$ and $C = \left(\frac{\pi}{2h}\right)^2 \left(c_l \alpha(T) + (g_l c_l - f_l^2)\left(\frac{\pi}{2h}\right)^2\right)$ are introduced. The solutions (15) can be essentially simplified in the case $c_l = c_\perp = c$, $g_l = g_\perp = g$ and $f_l = f_\perp = f$ corresponding to cubic m3m paraelectric (see, e.g., **Table I** for SrTiO$_3$) allowing for the hidden permutation symmetry of the static flexocoupling [37]. For the case the inverse static dielectric susceptibility

$$\frac{1}{\widetilde{\chi}_S(h, k_\perp)} = \alpha(T) + \left(g - \frac{f^2}{c}\right)\left(\left(\frac{\pi}{2h}\right)^2 + k_\perp^2\right) \quad (16)$$

tends to zero at

$$k_\perp^{cr}(T,h) = \frac{1}{R_C(T)} \sqrt{\frac{1}{(f^2/gc)-1} - \left(\frac{\pi R_C(T)}{2h}\right)^2}. \qquad (17)$$

Hereinafter the correlation radius $R_C(T) = \sqrt{\frac{g}{\alpha(T)}}$ is introduced. Since we consider the case $\alpha(T) > 0$, the solution (17) exists and so the SMP can occur when the absolute value of the static flexoelectric coupling constant is within the range

$$f_{cr}^{\min} < |f| < f_{cr}^{\max}(T,h). \qquad (18a)$$

The ends of the interval (18a) can be considered as the "minimal" and "maximal" critical values of the static flexoconstant, respectively. They are

$$f_{cr}^{\min} = \sqrt{gc}, \qquad f_{cr}^{\max}(T,h) = f_{cr}^{\min} \sqrt{1 + \left(\frac{2h}{\pi R_C(T)}\right)^2}. \qquad (18b)$$

Notably that the value $f_{cr}^{\min}$ obtained by us for $v=0$ is in agreement with Yudin et al [39] criteria obtained for the stability of homogeneous bulk material. However, the value $f_{cr}^{\max}(T,h)$ is a temperature and thickness-dependent parameter specific for thin ferroic films; it tends to infinity with the thickness increase.

If the flexconstant of the film is within the range $f_{cr}^{\min} < |f| < f_{cr}^{\max}$, the SMP with a period $k_\perp^{cr}(T,h)$ can occur in the film. The dependence of the dimensionless wave vector $k_\perp^{cr} R_C$ on the flexoconstant $|f|/\sqrt{gc}$ was calculated from Eq.(17) for several film thickness $2h = (8-80) R_C$ [see **Fig.4(a)**]. The value of $k_\perp^{cr}$ diverges at $f \to f_{cr}^{\min}$, decreases with $f$ increase and tends to zero at $f = f_{cr}^{\max}$. The dependence of the ratio $f_{cr}^{\max}/\sqrt{gc}$ on the dimensionless thickness $2h/R_C$ is presented in **Fig.4(b).** As shown in the figure, the ratio $f_{cr}^{\max}/\sqrt{gc}$ increases monotonically with $h$ increase. The dependence is sub-linear at $2h/R_C \leq 1$ and becomes linear ($f_{cr}^{\max} \sim h$) at $2h/R_C \gg 1$. When the flexoelectric coefficient is a known constant the value of $k_\perp^{cr}$ depends on the film thickness, as shown in **Fig. 4(c)**. As one can see, the value $k_\perp^{cr}$ appears at the critical thickness $2h = H_{cr}^f$, rapidly increases with $h$ increase and then saturates to the "bulk" value $k_\perp^{cr}(T,\infty) = \frac{1}{R_C(T)} \sqrt{\frac{1}{(f^2/gc)-1}}$. The critical thickness $H_{cr}^f$ can be derived from Eq.(17) and is equal to

$$H_{cr}^f(T) = \pi R_C(T)\sqrt{\frac{f^2}{gc}-1}. \qquad (19)$$

The dependence of the dimensionless ratio $H_{cr}^f/R_C$ on the flexoconstant $|f|/\sqrt{gc}$ is shown in **Fig. 4(d)**. The ratio critical thickness appears at $f = f_{cr}^{min}$ and monotonically increases with $f$ increase. In fact, **Fig. 4(d)** is a 90-degree rotated **Fig. 4(c)**, as it should be.

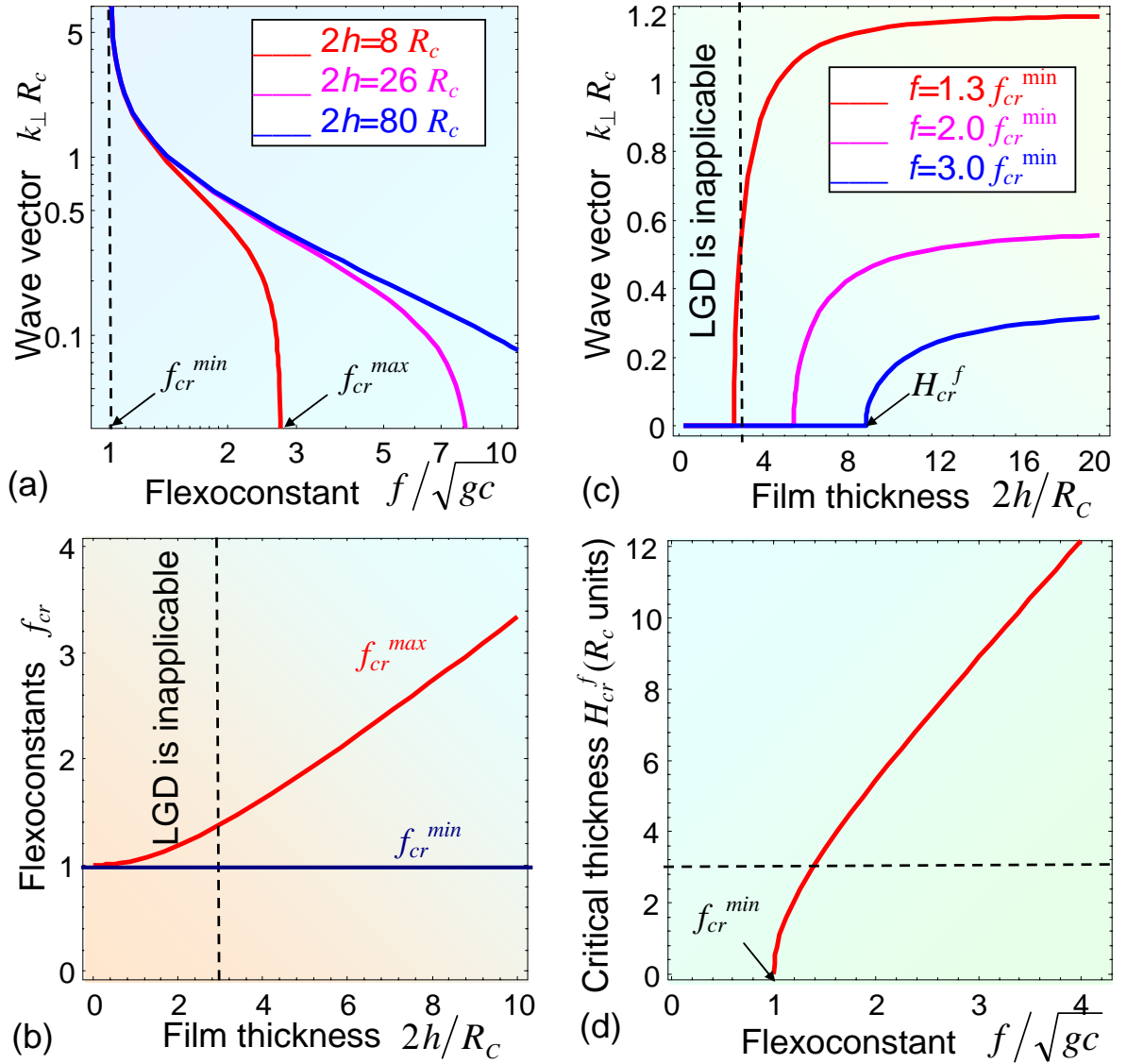

**FIG. 4.** (a) The dependence of $k_\perp^{cr}$ (in $R_C$ units) on the dimensionless flexoconstant $|f|/\sqrt{gc}$ calculated for several film thicknesses $2h = 8$, 26 and 80 $R_C$ (red, magenta and blue curves). (b) The dependences of the dimensionless critical flexoconstants $f_{cr}^{max}/\sqrt{gc}$ (red curve) and $f_{cr}^{min}/\sqrt{gc}$ (blue horizontal line) on the dimensionless thickness $2h/R_C$. (c) The dependence of $k_\perp^{cr}$ (in $R_C$ units) on the dimensionless ratio $2h/R_C$ calculated for several flexoconstants $|f|/\sqrt{gc} = 1.3$, 2 and 3 (red, magenta and blue curves). (d)

The dependence of the dimensionless critical thickness $H_{cr}^f/R_C$ on the flexoconstant $|f|/\sqrt{gc}$. Dashed vertical lines in plots (b) and (c) indicate the thickness limit of LGD approach applicability.

Note that all thickness dependences shown in **Figs. 4** can be considered as "true" size effect of the corresponding physical quantities. **Figures 4(a)-(d)** are not material specific, because they are plotted in dimensionless variables, $|f|/\sqrt{gc}$, $k_\perp^{cr} R_C$ and $2h/R_C$. For the considered case of SrTiO$_3$ the correlation radius $R_C$ is about 0.7 nm at 300 K, and the experimentally measured flexoelectric coefficient $f$ is about 2 V, is much smaller than the minimal critical value $f_{cr}^{\min} = 5.0$ V. Thus, there is no possibility to observe SMPs in SrTiO$_3$ thin films, since $f \ll f_{cr}^{\min}$, however it can be induced by flexoelectric effect in other ferroic films with higher flexoconstants (5 – 10)V. Note that the analytical expressions (17)-(19) can describe the influence of finite size effect on the appearance and properties of incommensurate spatial modulation in ferroic thin films.

## IV. CONCLUSION

Using Landau-Ginsburg-Devonshire theory and one-dimensional approximation describing two lowest transverse phonon modes, we derive and analyze analytical expressions for the frequency dispersion $\omega_{1,2}(\mathbf{k})$ of soft phonon modes in nanoferroics and perform numerical calculations for a thin SrTiO$_3$ film of thickness $2h$.

The frequency $\omega_1(h, k_\perp)$ monotonically decreases with $h$ increase, but it is nonzero at $k_\perp = 0$ for any thickness and tends to the expression [57] for the "bulk" acoustic mode only at $h \to \infty$. The result illustrates a reminiscent difference between the phonon spectra in thin film with discrete $k_z$ values and the spectra of bulk material with continuous $k_z$ values. The frequency $\omega_2(h, k_\perp)$ also monotonically decreases with $h$ increase and tends to the expression [57] for the "bulk" optic mode at $h \to \infty$.

The pronounced size effect (i.e. the minimum) occurs at the dependence of $\omega_2 - \omega_1$ on the film thickness $h$; and the minimum are localized in the range (6 – 20) nm for SrTiO$_3$ film. However it is questionable whether the prediction can be verified experimentally, because it is may be impossible to determine experimentally the phonon dispersion law at very small **k** in ultrathin films by inelastic neutron scattering, but probably possible by Raman or Brillouin scattering.

We derived analytical expressions describing the influence of the size effect on the appearance and properties of incommensurate spatial modulation in ferroic thin films. We revealed that SMP can exist in a thin film when the absolute value of the static flexoelectric coupling constant $|f|$ is within the range $f_{cr}^{\min} < |f| < f_{cr}^{\max}(T,h)$. It turned out that there is no possibility to observe SMPs in SrTiO$_3$ thin films, since corresponding flexoconstant $|f| \sim 2$ V is significantly smaller than the minimal critical value $f_{cr}^{\min} \sim 5$ V. However SMP can be induced by flexoelectric effect in other paraelectric films with higher flexoconstants, and the prediction can be verified experimentally by dielectric measurements and X-ray diffraction.

We revealed the pronounced "true" size effect in the dependence of soft phonon dispersion on the film thickness that can lead to the "apparent" or "false" size effect of dynamic flexoelectric coupling constants. The true and apparent size effects should be clearly distinguished from each other by experiment. However, it is not excluded that the origin of the "giant" flexoelectric coefficients extracted by several authors based on experimental results for a spatially confined or strongly inhomogeneous samples is the apparent size effect of dynamic constant. To verify the hypothesis the measurements of the soft phonon dispersion in nanoferroics seem urgent.

**Acknowledgements.** A.N.M. formulated the problem, performed analytical calculations and wrote the manuscript. E.A.E. performed numerical calculations. Both authors contributed equally to the discussion of obtained results. A.N.M. work was supported by the National Academy of Sciences of Ukraine (projects No. 0117U002612 and No. 0118U003535), and by the European Union's Horizon 2020 research and innovation programme under the Marie Skłodowska-Curie grant agreement No 778070.

## REFERENCES

[1] M.D. Glinchuk, A. V. Ragulya, V.A. Stephanovich, Nanoferroics, Springer (2013), p.378

[2] S. V. Kalinin, Y. Kim, D.D. Fong, and A. Morozovska. Surface Screening Mechanisms in Ferroelectric Thin Films and its Effect on Polarization Dynamics and Domain Structures. Rep. Prog. Phys. **81,** 036502 (2018)

[3] W.Cochran, Crystal stability and the theory of ferroelectricity, Phys. Rev. Lett. **3**, 412 (1959).

[4] G. Shirane, J. D. Axe, J. Harada, and J. P. Remeika. Soft ferroelectric modes in lead titanate, Phys. Rev. **B 2**, 155 (1970).

[5] W. Cochran, Dynamical, scattering and dielectric properties of ferroelectric crystals, Advances in Physics, **18**, 157 (1969).

[6] G. Shirane and Y. Yamada, Lattice-Dynamical Study of the 110 K Phase Transition in SrTiO$_3$, Phys. Rev., **177**, 858 (1969).

[7] R. Currat, H. Buhay, C. H. Perry, and A. M. Quittet, Inelastic neutron scattering study of anharmonic interactions in orthorhombic KNbO$_3$, Phys. Rev. **B 40**, 10741 (1989).

[8] I. Etxebarria, M. Quilichini, J. M. Perez-Mato, P. Boutrouille, F. J. Zuniga, and T. Breczewski, Inelastic neutron scattering investigation of external modes in incommensurate and commensurate $A_2BX_4$ materials, J. Phys.: Condensed Matter **4**, 8551 (1992).

[9] J. Hlinka, M. Quilichini, R. Currat, and J. F. Legrand, Dynamical properties of the normal phase of betaine calcium chloride dihydrate. I. Experimental results, J. Phys.: Condensed Matter **8**, 8207 (1996).

[10] J. Hlinka, S. Kamba, J. Petzelt, J. Kulda, C. A. Randall, and S. J. Zhang, Origin of the "Waterfall" effect in phonon dispersion of relaxor perovskites, Phys. Rev. Lett. **91**, 107602 (2003).

[11] V. Goian, S. Kamba, O. Pacherova, J. Drahokoupil, L. Palatinus, M. Dusek, J. Rohlıcek, M. Savinov, F. Laufek, W. Schranz, A. Fuith, M. Kachlık, K. Maca, A. Shkabko, L. Sagarna, A. Weidenkaff, and A. A. Belik, Antiferrodistortive phase transition in $EuTiO_3$, Phys. Rev. **B 86,** 054112 (2012).

[12] Jong-Woo Kim, P. Thompson, S.Brown, P. S. Normile, J. A. Schlueter, A. Shkabko, A. Weidenkaff, and P. J. Ryan, Emergent Superstructural Dynamic Order due to Competing Antiferroelectric and Antiferrodistortive Instabilities in Bulk $EuTiO_3$, Phys. Rev. Lett. **110**, 027201 (2013).

[13] R.G. Burkovsky, A. K. Tagantsev, K. Vaideeswaran, N. Setter, S. B. Vakhrushev, A. V. Filimonov, A. Shaganov, D. Andronikova, A. I. Rudskoy, A. Q. R. Baron, H. Uchiyama, D. Chernyshov, Z. Ujma, K. Roleder, A. Majchrowski, and Jae-Hyeon Ko, Lattice dynamics and antiferroelectricity in PbZrO 3 tested by x-ray and Brillouin light scattering, Phys. Rev. **B 90**, 144301 (2014).

[14] J. Hlinka, I. Gregora, and V. Vorlıcek, Complete spectrum of long-wavelength phonon modes in Sn2P2S6 by Raman scattering, Phys.Rev. **B 65**, 064308 (2002).

[15] A. Kohutych, R. Yevych, S. Perechinskii, V. Samulionis, J. Banys, and Yu Vysochanskii, Sound behavior near the Lifshitz point in proper ferroelectrics, Phys. Rev. **B 82**, no. 5: 054101 (2010).

[16] A. Kohutych, R. Yevych, S. Perechinskii, and Y. Vysochanskii, Acoustic attenuation in ferroelectric $Sn_2P_2S_6$ crystals, Open Phys. **8**, 905-914 (2010).

[17] Yu M. Vysochanskii, A. A. Kohutych, A. V. Kityk, A. V. Zadorozhna, M. M. Khoma, and A. A. Grabar, Tricritical Behavior of $Sn_2P_2S_6$ Ferroelectrics at Hydrostatic Pressure, Ferroelectrics 399, 83-88 (2010).

[18] R. M. Yevych, Yu. M. Vysochanskii, M. M. Khoma and S. I. Perechinskii, Lattice instability at phase transitions near the Lifshitz point in proper monoclinic ferroelectrics, J. Phys.: Condens. Matter **18**, 4047–4064 (2006).

[19] Flexoelectricity in Solids: From Theory to Applications, Ed. by A.K. Tagantsev and P.V. Yudin, World Scientific (2016).

[20] A.S. Yurkov, A. Dejneka, and P.V. Yudin, Flexoelectric polarization induced by inhomogeneous heating and implications for energy harvesting, Int. J. Solids and Structures (2018). https://doi.org/10.1016/j.ijsolstr.2018.12.003

[21] Sergei V. Kalinin and Anna N. Morozovska, Multiferroics: Focusing the light on flexoelectricity (Comment), Nat. Nanotechnol. 10, 916 (2015). doi:10.1038/nnano.2015.213

[22] A. N. Morozovska, V. V. Khist, M. D. Glinchuk, C. M. Scherbakov, M. V. Silibin, D. V. Karpinsky, and E. A. Eliseev. Flexoelectricity induced spatially modulated phases in ferroics and liquid crystals. (Author review) Journal of Molecular Liquids **267**, 550–559 (2018)

[23] A. K. Tagantsev, L. Eric Cross, and J. Fousek. Domains in Ferroic Crystals and Thin Films. (Springer New York, 2010)

[24] G. Catalan, L.J. Sinnamon and J.M. Gregg The effect of flexoelectricity on the dielectric properties of inhomogeneously strained ferroelectric thin films J. Phys.: Condens. Matter 16, 2253 (2004).

[25] M. S.Majdoub, R. Maranganti, and P. Sharma, Understanding the origins of the intrinsic dead layer effect in nanocapacitors, Phys. Rev. B 79, 115412 (2009).

[26] M. S. Majdoub, P. Sharma, and T. Cagin, Enhanced size-dependent piezoelectricity and elasticity in nanostructures due to the flexoelectric effect, Phys. Rev B **77**, 125424 (2008).

[27] R. Maranganti, and P. Sharma, Atomistic determination of flexoelectric properties of crystalline dielectrics. Phys. Rev. B 80, 054109 (2009).

[28] G. Catalan, A. Lubk, A. H. G. Vlooswijk, E. Snoeck, C. Magen, A. Janssens, G. Rispens, G. Rijnders, D. H. A. Blank and B. Noheda, Flexoelectric rotation of polarization in ferroelectric thin films, Nat. Mater. **10**, 963-967 (2011).

29 V. S. Mashkevich and K. B. Tolpygo, The interaction of vibrations of nonpolar crystals with electric fields, Zh. Eksp. Teor. Fiz. **31**, 520 (1957).

30 Sh. M. Kogan. Piezoelectric effect under an inhomogeneous strain and an acoustic scattering of carriers of current in crystals, Solid State Phys., **5**, 2829 (1963).

31 A.K. Tagantsev, Piezoelectricity and flexoelectricity in crystalline dielectrics, Phys. Rev B, **34**, 5883 (1986).

[32] P. Zubko, G. Catalan, A.K. Tagantsev, Flexoelectric Effect in Solids, Ann. Rev. of Mater. Res. 43, 387-421. (2013).

[33] P V Yudin and A K Tagantsev. Fundamentals of flexoelectricity in solids. Nanotechnology, 24, 432001 (2013).

[34] A. Kvasov, and A. K. Tagantsev, Dynamic flexoelectric effect in perovskites from first-principles calculations, Phys. Rev. B **92**, 054104 (2015).

[35] Longlong Shu, Xiaoyong Wei, Ting Pang, Xi Yao, and Chunlei Wang, Symmetry of flexoelectric coefficients in crystalline medium, J. Appl. Phys. 110, 104106 (2011); doi: 10.1063/1.3662196

[36] H. Le Quang, and Q-C. He, The number and types of all possible rotational symmetries for flexoelectric tensors, In Proceedings of the Royal Society of London A: Mathematical, Physical and Engineering Sciences, 467, 2369 (2011).

[37] Eugene A. Eliseev and Anna N. Morozovska, Hidden Symmetry of Flexoelectric Coupling, Phys. Rev. **B** 98, 094108 (2018).

[38] Alberto Biancoli, Chris M. Fancher, Jacob L. Jones, and Dragan Damjanovic. Breaking of macroscopic centric symmetry in paraelectric phases of ferroelectric materials and implications for flexoelectricity, Nat. Mater. **14**, 224 (2015).

[39] P. V. Yudin, R. Ahluwalia, A. K. Tagantsev, Upper bounds for flexocoupling coefficients in ferroelectrics, Appl. Phys. Lett. **104**, 082913 (2014).

[40] Jiawang Hong and D. Vanderbilt, First-principles theory of frozen-ion flexoelectricity, Phys. Rev. **B 84**, 180101(R) (2011).

[41] I. Ponomareva, A. K. Tagantsev, L. Bellaiche. Finite-temperature flexoelectricity in ferroelectric thin films from first principles, Phys. Rev. **B 85**, 104101 (2012).

[42] Jiawang Hong and D. Vanderbilt, First-principles theory and calculation of flexoelectricity, Phys. Rev. B **88**, 174107 (2013).

[43] M. Stengel, Flexoelectricity from density-functional perturbation theory, Phys. Rev. B **88**, 174106 (2013).

[44] M. Stengel, Unified ab initio formulation of flexoelectricity and strain-gradient elasticity, Phys. Rev. B 93, 245107 (2016).

[45] W. Ma, L.E. Cross, Strain-gradient-induced electric polarization in lead zirconate titanate ceramics, Appl. Phys. Lett., **82**, 3293 (2003).

[46] W. Ma and L. E. Cross, Flexoelectricity of barium titanate. Appl. Phys. Lett., **88**, 232902 (2006).

[47] W. Ma and L. E. Cross, Flexoelectric effect in ceramic lead zirconate titanate. Appl. Phys. Lett. **86**, 072905 (2005).

[48] D. Lee, A. Yoon, S.Y. Jang, J.-G. Yoon, J.-S. Chung, M. Kim, J. F. Scott, and T.W. Noh, Giant Flexoelectric Effect in Ferroelectric Epitaxial Thin Films, Phys. Rev. Lett. **107**, 057602 (2011).

[49] J. Lu, X. Liang, W. Yu, S. Hu, S. Shen, Temperature dependence of flexoelectric coefficient for bulk polymer polyvinylidene fluoride, J. Phys. D: Appl. Phys. Accepted manuscript https://doi.org/10.1088/1361-6463/aaf543

[50] X Wen, D Li, K Tan, Q Deng, S Shen, Flexoelectret: An Electret with Tunable Flexoelectric-like Response, arXiv:1811.09566.

[51] P. Zubko, G. Catalan, A. Buckley, P.R. L. Welche, J. F. Scott. Strain-gradient-induced polarization in $SrTiO_3$ single crystals, Phys. Rev. Lett. **99**, 167601 (2007).

[52] P. Zubko, G. Catalan, A. Buckley, P. R. L. Welche, and J. F. Scott, Erratum: Strain-Gradient-Induced Polarization in SrTiO3 Single Crystals [Phys. Rev. Lett. 99, 167601 (2007)], Phys. Rev. Lett. 100, 199906 (2008).

[53] L. P. Liu, and P. Sharma, Flexoelectricity and thermal fluctuations of lipid bilayer membranes: Renormalization of flexoelectric, dielectric, and elastic properties, Phys. Rev. E 87, 032715 (2013).

[54] Fatemeh Ahmadpoor, and Pradeep Sharma, Flexoelectricity in two-dimensional crystalline and biological membranes, Nanoscale 7, 16555-16570 (2015).

[55] I. B. Bersuker. Pseudo Jahn–Teller effect in the origin of enhanced flexoelectricity. Appl. Phys. Lett. 106, 022903 (2015); doi: 10.1063/1.4905679

[56] Anna N. Morozovska, Yulian M. Vysochanskii, Oleksandr V. Varenyk, Maxim V. Silibin, Sergei V. Kalinin, and Eugene A. Eliseev. Flexocoupling impact on the generalized susceptibility and soft phonon modes in the ordered phase of ferroics, Phys. Rev. B 92, 094308 (2015).

[57] Anna N. Morozovska, Eugene A. Eliseev, Christian M. Scherbakov, and Yulian M. Vysochanskii, The influence of elastic strain gradient on the upper limit of flexocoupling strength, spatially-modulated phases and soft phonon dispersion in ferroics, Phys. Rev. B 94, 174112 (2016).

[58] Anna N. Morozovska, Maya D. Glinchuk, Eugene A. Eliseev, and Yulian M. Vysochanskii. Flexocoupling-induced soft acoustic mode and the spatially modulated phases in ferroelectrics, Phys. Rev. B, 96, 094111 (2017).

[59] Anna N. Morozovska, Christian M. Scherbakov, and M.D. Glinchuk, Dependence of the soft phonon spectra on flexoelectric coupling in ferroelectrics, Ukr. J. Phys. 63, 168 (2018).

[60] Eugene A. Eliseev, Anna N. Morozovska, Maya D. Glinchuk, and Sergei V. Kalinin, Missed surface waves in non-piezoelectric solids, Phys. Rev. B 96, 045411 (2017).

61 Supplementary Materials [URL will be provided by Publisher]

62 S. W. H. Eijt, R. Currat, J. E. Lorenzo, P. Saint-Gregoire, B. Hennion, and Yu M. Vysochanskii, Soft modes and phonon interactions in Sn2P2S6 studied by neutron scattering, Euro. Phys. J. B **5**, 169-178 (1998).

63 S. W. H. Eijt, R. Currat, J. E. Lorenzo, P. Saint-Gregoire, S. Katano, T. Janssen, B. Hennion, and Yu M. Vysochanskii, Soft modes and phonon interactions in Sn2P2Se6 studied by means of neutron scattering, J. Phys.: Cond. Matter **10**, 4811 (1998).

64 C. Kappler, and M. B. Walker, Symmetry-based model for the modulated phases of betaine calcium chloride dihydrate, Phys. Rev. B 48, 5902 (1993).

65 J. Hlinka, M. Quilichini, R. Currat, and J. F. Legrand, Dynamical properties of the normal phase of betaine calcium chloride dihydrate. II. A semimicroscopic model, J. Phys.: Cond. Matter 8, 8221 (1996).

66. L.D. Landau and E.M. Lifshitz, Theory of Elasticity. Theoretical Physics, Vol. 7, Butterworth-Heinemann, Oxford, U.K. (1998).

67 G.A. Smolenskii, V.A. Bokov, V.A. Isupov, N.N Krainik, R.E. Pasynkov, A.I. Sokolov, Ferroelectrics and Related Materials (Gordon and Breach, New York, 1984).

68 J.H.Barrett, Dielectric Constant in Perovskite Type Crystals, Phys. Rev. 86, 118-120 (1952).

[69] For a geometry shown in Fig.1, assuming m3m symmetry of the paraelectric film and using the hidden permutative symmetry of the flexo-coupling one obtains that $f_\perp = f_l = f_{44} = f_{12}$

[70] E.A. Eliseev, A.N. Morozovska, M.D. Glinchuk, and R. Blinc. Spontaneous flexoelectric/flexomagnetic effect in nanoferroics, Phys. Rev. B. **79**, 165433 (2009).

71 A. S.Yurkov, Elastic boundary conditions in the presence of the flexoelectric effect. JETP Lett. **94**, 455-458 (2011).

72 Sheng Mao, and Prashant K. Purohit, Insights into flexoelectric solids from strain-gradient elasticity, J. Appl. Mech. **81**, 081004 (2014).

Supplementary Materials to

Size Effect of Soft Phonon Dispersion in Nanosized Ferroics

Anna N. Morozovska[1,2*] and Eugene A. Eliseev[3†]

[1] *Institute of Physics, National Academy of Sciences of Ukraine,*

*46, Prospekt Nauky, 03028 Kyiv, Ukraine*

[2] *Bogolyubov Institute for Theoretical Physics, National Academy of Sciences of Ukraine,*

*14-b Metrolohichna, 03680 Kyiv, Ukraine*

[3] *Institute for Problems of Materials Science, National Academy of Sciences of Ukraine,*

*3, Krjijanovskogo, 03142 Kyiv, Ukraine*

**APPENDIX**

Euler-Lagrange equations for $\eta(x,z)$ and $U(x,z)$, obtained from the variation of the free energy Eq.(1a), have the form:

$$\mu\frac{\partial^2\eta}{\partial t^2}+\alpha\eta+\beta\eta^3+\gamma\eta^5-g_l\frac{\partial^2\eta}{\partial z^2}-g_\perp\frac{\partial^2\eta}{\partial x^2}-f_l\frac{\partial^2 U}{\partial z^2}-f_\perp\frac{\partial^2 U}{\partial x^2}-2q\eta\frac{\partial U}{\partial z}+M\frac{\partial^2 U}{\partial t^2}=E, \quad (A.1)$$

$$v\frac{\partial^4 U}{\partial z^4}+\rho\frac{\partial^2 U}{\partial t^2}-c_l\frac{\partial^2 U}{\partial z^2}-c_\perp\frac{\partial^2 U}{\partial x^2}-f_l\frac{\partial^2\eta}{\partial z^2}-f_\perp\frac{\partial^2\eta}{\partial x^2}+2q\eta\frac{\partial\eta}{\partial z}+M\frac{\partial^2\eta}{\partial t^2}=-\vartheta\frac{\partial\delta N}{\partial z}. \quad (A.2)$$

The boundary conditions for $\eta(x,z)$, derived from the variation of the free energy Eqs.(1) on $\eta$, have the form:

$$\left.-\xi\eta-g_l\frac{\partial\eta}{\partial z}-\frac{f_l}{2}\left(\frac{\partial U}{\partial z}\right)\right|_{z=h}=0, \quad \left.\xi\eta-g_l\frac{\partial\eta}{\partial z}-\frac{f_l}{2}\left(\frac{\partial U}{\partial z}\right)\right|_{z=-h}=0. \quad (A.3a)$$

Supplementary boundary conditions for $U(x,z)$ in the considered geometry have the form

$$\left.\left(c_l\frac{\partial U}{\partial z}+f_l\frac{\partial\eta}{\partial z}-q\eta^2-v\frac{\partial^3 U}{\partial z^3}\right)\right|_{z=h}=0, \quad \left.\left(c_l\frac{\partial U}{\partial z}+f_l\frac{\partial\eta}{\partial z}-q\eta^2-v\frac{\partial^3 U}{\partial z^3}\right)\right|_{z=-h}=0. \quad (A.3b)$$

$$\left.\left(-\frac{f_l}{2}\eta+v\frac{\partial^2 U}{\partial z^2}\right)\right|_{z=h}=0, \quad \left.\left(-\frac{f_l}{2}\eta+v\frac{\partial^2 U}{\partial z^2}\right)\right|_{z=-h}=0. \quad (A.3c)$$

Let us seek for the solution of the boundary problem (A.2)-(A.3) in the paraelectric phase, where the spontaneous order parameter is zero. In Fourier k-ω domain, the integral expansions of η and $U$ on z-and x coordinates have the form:

$$\eta=\int dk_\perp dk_l \exp(ik_\perp x+ik_l z+i\omega t)\tilde\eta(\vec k), \quad U=\int dk_\perp dk_l \exp(ik_\perp x+ik_l z+i\omega t)\tilde U(\vec k) \quad (A.4)$$

[*] anna.n.morozovska@gmail.com
[†] eugene.a.eliseev@gmail.com

In Fourier space linearized Eqs. (A.1)-(A.2) acquire the form

$$\left(-\mu\omega^2 + \alpha + g_\perp k_\perp^2 + g_l k_l^2\right)\tilde{\eta} + \left(f_l k_l^2 + f_\perp k_\perp^2 - M\omega^2\right)\tilde{U} = \tilde{E}, \qquad (A.5a)$$

$$\left(vk_l^4 + c_\perp k_\perp^2 + c_l k_l^2 - \rho\omega^2\right)\tilde{U} + \left(f_l k_l^2 + f_\perp k_\perp^2 - M\omega^2\right)\tilde{\eta} = -ik_l \tilde{N}. \qquad (A.5b)$$

As a result the Fourier images of the order parameter $\tilde{\eta}(\vec{k})$ and elastic displacement $\tilde{U}(\vec{k})$ are linearly proportional to external electric field and mechanical force variations, $\tilde{E}$ and $\tilde{N} = -ik_l N$. The formal solution of Eqs.(A.5) can be presented in the matrix form

$$\begin{pmatrix}\tilde{\eta}\\ \tilde{U}\end{pmatrix} = \Xi \begin{pmatrix}\tilde{E}\\ \tilde{N}\end{pmatrix}, \qquad \Xi = \begin{pmatrix}\tilde{\chi}(\vec{k},\omega) & \tilde{\xi}(\vec{k},\omega)\\ \tilde{\xi}^*(\vec{k},\omega) & \tilde{\vartheta}(\vec{k},\omega)\end{pmatrix} \qquad (A.6)$$

where the elements of matrix $\Xi$, which are generalized susceptibilities, are given by expressions:

$$\tilde{\chi}(\vec{k},\omega) = \frac{vk_l^4 + c_\perp k_\perp^2 + c_l k_l^2 - \rho\omega^2}{\Delta(\vec{k},\omega)}, \qquad \tilde{\xi}(\vec{k},\omega) = -\frac{f_l k_l^2 + f_\perp k_\perp^2 - M\omega^2}{\Delta(\vec{k},\omega)}, \qquad (A.7a)$$

$$\tilde{\vartheta}(\vec{k},\omega) = \frac{\alpha + g_\perp k_\perp^2 + g_l k_l^2 - \mu\omega^2}{\Delta(\vec{k},\omega)}, \qquad (A.7b)$$

$$\Delta(\vec{k},\omega) = \left(\alpha + g_l k_l^2 + g_\perp k_\perp^2 - \mu\omega^2\right)\left(vk_l^4 + ck_l^2 + c_\perp k_\perp^2 - \rho\omega^2\right) - \left(f_l k_l^2 + f_\perp k_\perp^2 - M\omega^2\right)^2. \qquad (A.7c)$$

The condition $\Delta(\vec{k},\omega) = 0$ gives us the biquadratic equation for the eigen frequency dispersion, $\omega(k)$, in a **bulk ferroic**, $\left(\mu\rho - M^2\right)\omega^4 - C(\vec{k})\omega^2 + B(\vec{k}) = 0$. The solution of the biquadratic equation can be represented in the form:

$$\omega_{1,2}^2(\vec{k}) = \frac{C(\vec{k}) \pm \sqrt{C^2(\vec{k}) - 4(\mu\rho - M^2)B(\vec{k})}}{2(\mu\rho - M^2)}, \qquad (A.8)$$

where the functions $C(\vec{k}) = \alpha\rho + (\mu c_\perp - 2f_\perp M + \rho g_\perp)k_\perp^2 + (c_l\mu - 2f_l M + g\rho)k_l^2 + \mu v k_l^4$ and $B(\vec{k}) = \left(\alpha + g_l k_l^2 + g_\perp k_\perp^2\right)\left(vk_l^4 + c_l k_l^2 + c_\perp k_\perp^2\right) - \left(f_l k_l^2 + f_\perp k_\perp^2\right)^2$, respectively. Dispersion relation (A.8) contains one optical (**O**) and one acoustic (**A**) phonon modes, which corresponds to the signs "+" and "−" before the radical, respectively. The **O**-mode is in fact transverse, and the **A**-mode can be longitudinal or transverse.

Since we are primary interested in a generalized susceptibility as a film response to external electric field, one can put $\tilde{N} = 0$ in Eq.(A.5b) and obtain the following relation between $\tilde{\eta}(\vec{k})$ and $\tilde{U}(\vec{k})$ spectra:

$$\tilde{U}(\vec{k},\omega) = -\frac{f_l k_l^2 + f_\perp k_\perp^2 - M\omega^2}{vk_l^4 + c_l k_l^2 + c_\perp k_\perp^2 - \rho\omega^2}\tilde{\eta}(\vec{k},\omega), \qquad \tilde{N} = 0 \qquad (A.9)$$

For thin films, the substitution of Equations (A.6) to the boundary conditions (A.3) makes the $k_l$-spectra of $\omega(\vec{k})$ discrete, $k_l \to k_m$, $m$ are integer numbers. On the other hand the condition $\Delta(\vec{k},\omega) = 0$ gives us the six order equation for k at fixed $\omega$ in a bulk ferroic. All 6 formal solutions, which can be complex numbers, further designed as $\pm k_m^{(1)}$, $\pm k_m^{(2)}$ and $\pm k_m^{(3)}$, should be accounted in thin films. As anticipated $\omega(\pm k_m^{(1)}) = \omega(\pm k_m^{(2)}) = \omega(\pm k_m^{(3)}) \equiv \omega(k_m)$

Since the eigen frequency depends on $k_m^2$ per equation $\Delta(k,\omega) = 0$, also valid for thin films, the general solution of homogeneous equations (A.5) acquire the form:

$$\eta = \sum_{n=1}^{3}\sum_{m=1}^{\infty}\int dk_\perp \left(\tilde{\eta}_{1m}^{(n)}\exp[ik_m^{(n)}z] + \tilde{\eta}_{2m}^{(n)}\exp[-ik_m^{(n)}z]\right)\exp[ik_\perp x + i\omega(k_m,k_\perp)t], \quad (A.10a)$$

$$U = \sum_{n=1}^{3}\sum_{m=1}^{\infty}\int dk_\perp \left(\tilde{U}_{1m}^{(n)}\exp[ik_m^{(n)}z] + \tilde{U}_{2m}^{(n)}\exp[-ik_m^{(n)}z]\right)\exp[ik_\perp x + i\omega(k_m,k_\perp)t]. \quad (A.10b)$$

Allowing for Eq.(A.9)

$$\tilde{U}_{1m}^{(n)} = -\frac{f_l(k_m^{(n)})^2 + f_\perp k_\perp^2 - M\omega^2}{v(k_m^{(n)})^4 + c_l(k_m^{(n)})^2 + c_\perp k_\perp^2 - \rho\omega^2}\tilde{\eta}_{1m}^{(n)}, \quad (A.10c)$$

$$\tilde{U}_{2m}^{(n)} = -\frac{fk_m^2 + f_\perp k_\perp^2 - M\omega^2}{v(k_m^{(n)})^4 + c_l(k_m^{(n)})^2 + c_\perp k_\perp^2 - \rho\omega^2}\tilde{\eta}_{2m}^{(n)}. \quad (A.10d)$$

Perturbations $\tilde{E}$ and $\tilde{N}$ spectra are continuous, and can be expanded in a full series of eigen functions. The boundary conditions (A.3) along with relations (A.9) give six equations for the six unknown spectral constants.

The remained conditions (A.3a), $\mp \zeta\eta - g_l\frac{\partial \eta}{\partial z} - \frac{f_l}{2}\left(\frac{\partial U}{\partial z}\right)\bigg|_{z=\pm h} = 0$, lead to the equations for the Fourier components:

$$\sum_{n=1}^{3}\left[\begin{array}{l}\left((-\zeta + ik_m^{(n)}g_l)\tilde{\eta}_{1m}^{(n)} - ik_m^{(n)}\frac{f_l}{2}\tilde{U}_{1m}^{(n)}\right)\exp(ik_m^{(n)}h) + \\ \left((-\zeta - ik_m^{(n)}g_l)\tilde{\eta}_{2m}^{(n)} + ik_m^{(n)}\frac{f_l}{2}\tilde{U}_{2m}^{(n)}\right)\exp(-ik_m^{(n)}h)\end{array}\right] = 0, \quad (A.11a)$$

$$\sum_{n=1}^{3}\left[\begin{array}{l}\left((\zeta + ik_m^{(n)}g_l)\tilde{\eta}_{1m}^{(n)} - ik_m^{(n)}\frac{f_l}{2}\tilde{U}_{1m}^{(n)}\right)\exp(-ik_m^{(n)}h) + \\ \left((\zeta - ik_m^{(n)}g_l)\tilde{\eta}_{2m}^{(n)} + ik_m^{(n)}\frac{f_l}{2}\tilde{U}_{2m}^{(n)}\right)\exp(ik_m^{(n)}h)\end{array}\right] = 0. \quad (A.11b)$$

The explicit form of conditions $\left(c_l\frac{\partial U}{\partial z} + f_l\frac{\partial \eta}{\partial z} - q\eta^2 - v\frac{\partial^3 U}{\partial z^3}\right)\bigg|_{z=\pm h} \approx \left(c_l\frac{\partial U}{\partial z} + f_l\frac{\partial \eta}{\partial z} - v\frac{\partial^3 U}{\partial z^3}\right)\bigg|_{z=\pm h} = 0$

is:

$$\sum_{n=1}^{3}\left[\begin{array}{l}\left(ik_m^{(n)}f_l\tilde{\eta}_{1m}^{(n)} + (ik_m^{(n)}c_l + iv(k_m^{(n)})^3)\tilde{U}_{1m}^{(n)}\right)\exp(ik_m^{(n)}h) + \\ \left(-ik_m^{(n)}f_l\tilde{\eta}_{1m}^{(n)} + (-ik_m^{(n)}c_l - iv(k_m^{(n)})^3)\tilde{U}_{1m}^{(n)}\right)\exp(-ik_m^{(n)}h)\end{array}\right] = 0, \quad (A.12a)$$

$$\sum_{n=1}^{3}\left[\begin{array}{c}\left(ik_m^{(n)} f \tilde{\eta}_{1m}^{(n)} + \left(ik_m^{(n)} c_l + iv\left(k_m^{(n)}\right)^3\right)\tilde{U}_{1m}^{(n)}\right)\exp\left(-ik_m^{(n)} h\right) + \\ \left(-ik_m^{(n)} f_l \tilde{\eta}_{1m}^{(n)} + \left(-ik_m^{(n)} c_l - iv\left(k_m^{(n)}\right)^3\right)\tilde{U}_{1m}^{(n)}\right)\exp\left(ik_m^{(n)} h\right)\end{array}\right] = 0. \quad \text{(A.12b)}$$

The explicit form of the conditions $\left(-\dfrac{f_l}{2}\eta + v\dfrac{\partial^2 U}{\partial z^2}\right)\bigg|_{z=\pm h} = 0$ is:

$$\sum_{n=1}^{3}\left[\left(-\frac{f_l}{2}\tilde{\eta}_{1m}^{(n)} - v\left(k_m^{(n)}\right)^2 \tilde{U}_{1m}^{(n)}\right)\exp\left(ik_m^{(n)} h\right) + \left(-\frac{f_l}{2}\tilde{\eta}_{2m}^{(n)} - v\left(k_m^{(n)}\right)^2 \tilde{U}_{2m}^{(n)}\right)\exp\left(-ik_m^{(n)} h\right)\right] = 0, \quad \text{(A.13a)}$$

$$\sum_{n=1}^{3}\left[\left(-\frac{f_l}{2}\tilde{\eta}_{1m}^{(n)} - v\left(k_m^{(n)}\right)^2 \tilde{U}_{1m}^{(n)}\right)\exp\left(-ik_m^{(n)} h\right) + \left(-\frac{f_l}{2}\tilde{\eta}_{2m}^{(n)} - v\left(k_m^{(n)}\right)^2 \tilde{U}_{2m}^{(n)}\right)\exp\left(ik_m^{(n)} h\right)\right] = 0. \quad \text{(A.13b)}$$

Equations (A.11)-(A.13) should be solved along with the relations between the order parameter $\tilde{\eta}(\vec{k})$ and elastic displacement $\tilde{U}(\vec{k})$, namely

$$\tilde{U}_{1m}^{(n)} + \frac{f_l\left(k_m^{(n)}\right)^2 + f_\perp k_\perp^2 - M\omega^2}{v\left(k_m^{(n)}\right)^4 + c_l\left(k_m^{(n)}\right)^2 + c_\perp k_\perp^2 - \rho\omega^2}\tilde{\eta}_{1m}^{(n)} = 0, \quad \text{(A.14a)}$$

$$\tilde{U}_{2m}^{(n)} + \frac{f_l\left(k_m^{(n)}\right)^2 + f_\perp k_\perp^2 - M\omega^2}{v\left(k_m^{(n)}\right)^4 + c_l\left(k_m^{(n)}\right)^2 + c_\perp k_\perp^2 - \rho\omega^2}\tilde{\eta}_{2m}^{(n)} = 0. \quad \text{(A.14b)}$$

Equations (A.11)-(A.14) have the very cumbersome 8x8 matrix form similar to Eq.(A.6), and can be solved numerically. Further calculations of derived determinant 6x6 are very cumbersome and unfortunately they do not allow closed form analytical results. The circumstances subjected us to look for approximate analytical solutions.

### B. Approximate solution for thin films

To obtain and analyze relatively simple analytical expressions, one can show that the conditions (A.3b) and (A.3c) can be neglected in the case of small enough $v$ and under the validity of strict equality $f^2 \ll cg$, regarded valid hereinafter. Notably, the conditions (A.3b) and (A.3c) become excessive in the limit $v \to 0$, considered hereinafter.

In the limit $v \to 0$ the remained conditions (A.3a), $\mp \zeta\eta - g_l\dfrac{\partial \eta}{\partial z} - \dfrac{f_l}{2}\left(\dfrac{\partial U}{\partial z}\right)\bigg|_{z=\pm h} = 0$, lead to the equations for the Fourier components:

$$\left(\tilde{\eta}_{m1}(ik_m g_l - \zeta) - ik_m\frac{f_l}{2}\tilde{U}_{m1}\right)\exp(ik_m h) + \left(-\tilde{\eta}_{m2}(\zeta + ik_m g_l) + ik_m\frac{f_l}{2}\tilde{U}_{m2}\right)\exp(-ik_m h) = 0, \quad \text{(A.15a)}$$

$$\left(\tilde{\eta}_{m1}(\zeta + ik_m g_l) - ik_m\frac{f_l}{2}\tilde{U}_{m1}\right)\exp(-ik_m h) + \left(\tilde{\eta}_{m2}(\zeta - ik_m g_l) + ik_m\frac{f_l}{2}\tilde{U}_{m2}\right)\exp(ik_m h) = 0. \quad \text{(A.15b)}$$

In Eqs.(A.15) we omitted the subscript *n*, since only two boundary conditions are supposed to be relevant. Allowing for the relation between $\tilde{\eta}(k)$ and $\tilde{U}(k)$ [see Eqs.(A.9c)], Eqs.(A.15) can be written in a matrix form

$$\begin{pmatrix} D_{11}^m & D_{12}^m \\ D_{21}^m & D_{22}^m \end{pmatrix} \begin{pmatrix} \tilde{\eta}_{m1} \\ \tilde{\eta}_{m2} \end{pmatrix} = 0, \quad (A.16)$$

where

$$D_{11}^m = \left( -\zeta + ik_m g_l + ik_m \frac{f_l}{2} \frac{f_l k_m^2 + f_\perp k_\perp^2 - M\omega^2}{c_l k_m^2 + c_\perp k_\perp^2 - \rho\omega^2} \right) \exp(ik_m h), \quad (A.17a)$$

$$D_{12}^m = \left( -\zeta - ik_m g_l - ik_m \frac{f_l}{2} \frac{f_l k_m^2 + f_\perp k_\perp^2 - M\omega^2}{c_l k_m^2 + c_\perp k_\perp^2 - \rho\omega^2} \right) \exp(-ik_m h), \quad (A.17b)$$

$$D_{21}^m = \left( \zeta + ik_m g_l + ik_m \frac{f_l}{2} \frac{f_l k_m^2 + f_\perp k_\perp^2 - M\omega^2}{c_l k_m^2 + c_\perp k_\perp^2 - \rho\omega^2} \right) \exp(-ik_m h), \quad (A.17c)$$

$$D_{22}^m = \left( \zeta - ik_m g_l - ik_m \frac{f_l}{2} \frac{f_l k_m^2 + f_\perp k_\perp^2 - M\omega^2}{c_l k_m^2 + c_\perp k_\perp^2 - \rho\omega^2} \right) \exp(ik_m h). \quad (A.17d)$$

Determinant of the system (A.16) has the form

$$\|D_m\| = D_{11}^m D_{22}^m - D_{12}^m D_{21}^m \equiv A(k_m, \omega)\exp(2ik_m h) - A^*(k_m, \omega)\exp(-2ik_m h). \quad (A.18a)$$

Where

$$A(k_m, \omega) = -\left( \zeta - ik_m g_l - ik_m \frac{f_l}{2} \frac{f_l k_m^2 + f_\perp k_\perp^2 - M\omega^2}{c_l k_m^2 + c_\perp k_\perp^2 - \rho\omega^2} \right)^2 \quad (A.18b)$$

The condition $\|D_m\| = 0$ is equivalent to the equation $\text{Im}[A(k_m, \omega)\exp(2ik_m h)] = 0$, or in another equivalent form

$$\text{Im}[A(k_m, \omega)]\cos(2k_m h) + \text{Re}[A(k_m, \omega)]\sin(2k_m h) = 0. \quad (A.19)$$

In the limit $v=0$ the expression for dynamic dielectric susceptibility to external electric field [given by Eq.(A.7a)] acquires the form

$$\tilde{\chi}(k_m, k_\perp, \omega) = \frac{1}{\alpha(T) + g_l k_m^2 + g_\perp k_\perp^2 - \mu\omega^2 - \dfrac{\left(f_l k_m^2 + f_\perp k_\perp^2 - M\omega^2\right)^2}{c_l k_m^2 + c_\perp k_\perp^2 - \rho\omega^2}}, \quad (A.20)$$

The dynamic dielectric susceptibility depends on the film thickness $h$ since $k_m$ should be *h*-dependent.

The soft phonons dispersion $\omega(k_m, k_\perp)$ is given by Eq.(A.8) at $v=0$, namely:

$$\omega_{1,2}^2(k_m, k_\perp) = \frac{C(k_m, k_\perp) \pm \sqrt{C^2(k_m, k_\perp) - 4(\mu\rho - M^2)B(k_m, k_\perp)}}{2(\mu\rho - M^2)}, \quad (A.21)$$

where the functions $C(k_m, k_\perp) = \alpha\rho + (\mu c_\perp - 2f_\perp M + \rho g_\perp)k_\perp^2 + (c_l\mu - 2f_l M + g\rho)k_m^2$ and $B(k_m, k_\perp) = (\alpha + g_l k_m^2 + g_\perp k_\perp^2)(c_l k_m^2 + c_\perp k_\perp^2) - (f_l k_m^2 + f_\perp k_\perp^2)^2$ also depend on the film thickness $h$ via the thickness dependence of $k_m$.

### C. Particular cases

**Case I. The flexocoupling is absent.** For the reference case of the flexocoupling absence, i.e. for $f_l = f_\perp = 0$ and $M = 0$, one get $A(k_m, \omega) = -(\zeta - ik_m g_l)^2$ in Eq.(A.18b). This leads to the characteristic equation

$$2\zeta k_m g_l \cos(2k_m h) + \left[(k_m g)^2 - \zeta^2\right]\sin(2k_m h) = 0 \qquad (A.22a)$$

Eq.(A.18a) is equivalent to the equation $\dfrac{2\zeta k_m g_l}{\zeta^2 - (k_m g)^2} = \tan(2k_m h)$ that is equivalent to the relations

$$\left[\begin{array}{l} \tan(k_m h) = \dfrac{k_m g_l}{\zeta}, \\ \tan(k_m h) = -\dfrac{\zeta}{k_m g_l}. \end{array}\right. \qquad (A.22b)$$

Note that the boundary conditions (A.3a) can be rewritten as $\left.\mp\eta - \dfrac{g_l}{\zeta}\dfrac{\partial\eta}{\partial z}\right|_{z=\pm h} = 0$ for the case of zero flexocoupling. The solution (A.22) is well-known (see e.g. Ref.[1]).

A limiting case $\zeta \to \infty$ corresponds to zero order parameter at the film surfaces, $\eta|_{z=\pm h} = 0$, and the solution of Eqs.(A.22) is $\tan(k_m h) \to 0$ or $\tan(k_m h) \to \pm\infty$, that gives $k_m = \dfrac{\pi m}{2h}$, where $m = 0, 1, 2,...$ Since the lowest root $m=0$ corresponds to constant solution for the order parameter, that does not satisfy the boundary condition $\eta|_{z=\pm h} = 0$, one should consider the first root $m=1$ corresponding to the first harmonic in the order parameter distribution consistent with the condition $\eta|_{z=\pm h} = 0$. Corresponding solution has the form:

$$\eta = \int dk_\perp \tilde\eta(k_\perp, k_l)\exp(ik_\perp x + i\omega t)\cos\left[\dfrac{\pi z}{2h}\right], \quad k_l = \dfrac{\pi}{2h}, \qquad (A.23)$$

For the case dynamic susceptibility given by Eq.(A.20) simplifies to the following expression:

$$\tilde\chi(k_\perp, \omega) = \dfrac{1}{\alpha(T) + g_l(\pi/2h)^2 + g_\perp k_\perp^2 - \mu\omega^2}. \qquad (A.24)$$

The soft phonons dispersion $\omega(h, \vec k)$ given by Eq.(A.17) acquires the form:

$$\omega_{1,2}^2(h,k_\perp) = \frac{C(h,k_\perp) \pm \sqrt{C^2(h,k_\perp) - 4(\mu\rho - M^2)B(h,k_\perp)}}{2(\mu\rho - M^2)}, \tag{A.25}$$

where the functions $C(h,k_\perp) \approx \alpha(T)\rho + (\mu c_\perp + \rho g_\perp)k_\perp^2 + (c_l\mu + g_l\rho)(\pi/2h)^2$ and $B(h,k_\perp) \approx [\alpha(T) + g_l(\pi/2h)^2 + g_\perp k_\perp^2] \cdot [c_l(\pi/2h)^2 + c_\perp k_\perp^2]$ depend on the film thickness $h$.

Note that the particular case $\zeta \to 0$ corresponds to zero flux of the order parameter at the film surfaces, $\left.\frac{\partial \eta}{\partial z}\right|_{z=\pm h} = 0$, and the solution of Eqs.(A.18) is again $k_m = \frac{\pi m}{2h}$, where $m = 0, 1, 2,...$ However for the case one should consider the lowest root $m=0$ corresponding to the constant order parameter, $\eta = \eta_S$, that gives the dispersion law and susceptibility as in the bulk sample.

**Case II. The flexocoupling is present.** For the particular case of the zero order parameter $\eta|_{z=\pm h} = 0$ at the film surfaces, the constant $\zeta \to \infty$ in the boundary conditions (A.3a), since the conditions should be rewritten as $\left.\mp\eta - \frac{g_l}{\zeta}\frac{\partial \eta}{\partial z} - \frac{f_l}{2\zeta}\left(\frac{\partial U}{\partial z}\right)\right|_{z=\pm h} = 0$. Since for $A(k_m,\omega) \to -\zeta^2$ for $\zeta \to \infty$, the condition $\sin(2k_m h) = 0$ should be valid in Eq.(A.14). The equation $\sin(2k_m h) = 0$ gives a well-known solution for eigen wave numbers $k_m = \frac{\pi m}{2h}$, where $m = 0, 1, 2,...$. Since $m=0$ corresponds to identically zero solution for the order parameter, one should consider $m=1$ corresponding to the lowest root. The first harmonic in the order parameter distribution is the same as Eq.(A.23),

$$\eta = \int dk_\perp \exp(ik_\perp x + i\omega t)\tilde{\eta}(k_\perp, k_l)\cos\left[\frac{\pi z}{2h}\right], \quad k_l = \frac{\pi}{2h}, \tag{A.26}$$

For the case dynamic susceptibility is given by the following expression:

$$\tilde{\chi}(k_\perp, \omega) = \frac{1}{\alpha(T) + g_l\left(\frac{\pi}{2h}\right)^2 + g_\perp k_\perp^2 - \mu\omega^2 - \frac{(f_l(\pi/2h)^2 + f_\perp k_\perp^2 - M\omega^2)^2}{c_l(\pi/2h)^2 + c_\perp k_\perp^2 - \rho\omega^2}}, \tag{A.27}$$

The soft phonons dispersion $\omega(h,\vec{k})$ given by Eq.(A.21) acquires the form:

$$\omega_{1,2}^2(h,k_\perp) = \frac{C(h,k_\perp) \pm \sqrt{C^2(h,k_\perp) - 4(\mu\rho - M^2)B(h,k_\perp)}}{2(\mu\rho - M^2)}, \tag{A.28}$$

where the functions $C(h,k_\perp) \approx \alpha(T)\rho + (\mu c_\perp - 2f_\perp M + \rho g_\perp)k_\perp^2 + (c_l\mu - 2f_l M + g_l\rho)(\pi/2h)^2$ and $B(h,k_\perp) \approx (\alpha(T) + g_l(\pi/2h)^2 + g_\perp k_\perp^2)(c_l(\pi/2h)^2 + c_\perp k_\perp^2) - (f_l(\pi/2h)^2 + f_\perp k_\perp^2)^2$ depend on the film thickness $h$.

Another limiting case is zero derivative of the order parameter $\left.\frac{\partial \eta}{\partial z}\right|_{z=\pm h} = 0$ at the film surfaces corresponding to $\zeta \to 0$. The condition $\zeta \to 0$ leads to the expression $A(k_m, \omega) = \left(g_l + \frac{f_l}{2} \frac{f_l k_m^2 + f_\perp k_\perp^2 - M\omega^2}{c_l k_m^2 + c_\perp k_\perp^2 - \rho\omega^2}\right)^2 k_m^2$ and so the solution of Eq.(A.19), that is again $\sin(2k_m h) = 0$, has the form $k_m = \frac{\pi m}{2h}$, where $m = 0, 1, 2, \ldots$. However for the case one should consider the lowest root $m=0$ corresponding to the constant order parameter and bulk expressions for generalized susceptibility and soft phonon spectra.

### C. Derivation of approximate expressions (12)-(13)

From Eq.(11), the expressions are valid

$$\omega_1^2(h,0) = \frac{2B(h,0)}{C(h,0) + \sqrt{C^2(h,0) - 4(\mu\rho - M^2)B(h,0)}} \approx \frac{B(h,0)}{C(h,0)}$$

$$\cong \frac{\alpha c + (g_l c_l - f_l^2)(\pi/2h)^2}{\alpha\rho + (c_l\mu - 2f_l M + g_l\rho)(\pi/2h)^2} \left(\frac{\pi}{2h}\right)^2 \approx \frac{c_l}{\rho}\left(\frac{\pi}{2h}\right)^2 + \frac{g_l c_l - f_l^2}{\alpha\rho}\left(\frac{\pi}{2h}\right)^4$$

(A.29a)

$$\omega_2^2(h,0) = \frac{C(h,0) + \sqrt{C^2(h,0) - 4(\mu\rho - M^2)B(h,0)}}{2(\mu\rho - M^2)} \approx \frac{C(h,0)}{\mu\rho - M^2}$$

$$\cong \frac{\alpha\rho}{\mu\rho - M^2} + \frac{c_l\mu - 2f_l M + g_l\rho}{\mu\rho - M^2}\left(\frac{\pi}{2h}\right)^2 \approx \frac{\alpha}{\mu} + \frac{c_l\mu - 2f_l M + g_l\rho}{\mu\rho}\left(\frac{\pi}{2h}\right)^2$$

(A.29b)

Approximate equality in Eqs.(A.29) is valid for enough high thicknesses $h$ and $\alpha > 0$.

Elementary transformations of Eqs.(11) lead to the quadratic equation for $M$:

$$\left[\alpha(T)\rho - (\omega_2^2 + \omega_1^2)(\mu\rho - M^2)\right] - 2f_l M\left(\frac{\pi}{2h}\right)^2 + (c_l\mu + g_l\rho)\left(\frac{\pi}{2h}\right)^2 = 0 \quad \text{(A.30)}$$

Elementary transformations of Eqs.(A.30) lead to the expression for the $M$:

$$M_{1,2} = f_l\left(\frac{\pi}{2h\omega}\right)^2 \pm \sqrt{f_l^2\left(\frac{\pi}{2h\omega}\right)^4 - \left(\frac{\pi}{2h\omega}\right)^2(c_l\mu + g_l\rho) + \rho\left(\mu - \frac{\alpha}{\omega^2}\right)}$$

$$\approx \begin{cases} \frac{1}{f_l}\left[\rho(\alpha - \omega^2\mu)\left(\frac{2h}{\pi}\right)^2 + c_l\mu + g_l\rho\right], & h \to 0 \\ \pm\sqrt{\rho\left(\mu - \frac{\alpha}{\omega^2}\right)}, & h \to \infty \end{cases} \quad \text{(A.31)}$$

### REFERENCES

1 E.A. Eliseev, A.N. Morozovska. General approach to the description of the size effect in ferroelectric nanosystems. J. Mater. Science. **44**, 5149-5160 (2009).